\documentclass[%
 reprint,
 superscriptaddress,
 amsmath,amssymb,
 aps,
 pra,
 longbibliography,
lengthcheck,%
]{revtex4-1}

\usepackage{graphicx}
\usepackage{dcolumn}
\usepackage{bm}
\usepackage{hyperref}
\usepackage{color}
\usepackage{ulem}
\usepackage{xr}
\begin{document}

\preprint{APS/123-QED}

\title{
  Robustness of anomaly-related magnetoresistance in doped Weyl semimetals
}

\author{Hiroaki Ishizuka}
\affiliation{
Department of Applied Physics, The University of Tokyo, Bunkyo, Tokyo, 113-8656, JAPAN 
}

\author{Naoto Nagaosa}
\affiliation{
Department of Applied Physics, The University of Tokyo, Bunkyo, Tokyo, 113-8656, JAPAN 
}
\affiliation{
RIKEN Center for Emergent Matter Sciences (CEMS), Wako, Saitama, 351-0198, JAPAN
}

\date{\today}

\begin{abstract}
Weyl semimetal with Weyl fermions at Fermi energy is one of the topological materials, and is a condensed-matter realization of the relativistic fermions. However, there are several crucial differences such as the shift of Fermi energy, which can hinder the expected interesting physics. Chiral anomaly is a representative nontrivial phenomenon associated with Weyl fermions, which dictates the transfer of fermions between the Weyl fermions with opposite chirality; it is manifested as the negative magnetoresistance. Here we demonstrate that the magnetoresistance is robust against the deviation from the ideal Weyl Hamiltonian such as the shifted Fermi energy and nonlinear dispersions. We study a model with the energy dispersion containing two Weyl nodes, and find that the magnetoresistance persists even when the Fermi level is far away from the node, even above the saddle point that separates the two nodes. Surprisingly, the magnetoresistance remains even after the pair annihilation of the nodes.
\end{abstract}

\pacs{
}

\maketitle

\section{Introduction}
Quantum anomaly, the violation of conservation laws by quantum effect, has a long history of study in the field of quantum field theory, deeply rooted in its foundation~\cite{Fujikawa2004}. One such phenomenon is the chiral anomaly, which was initially discovered in the problem of photon self-energy~\cite{Fujikawa2004,Fukuda1949a,Fukuda1949b,Fukuda1949c,Steinberger1949}. Later, the violation of the chiral symmetry was also discovered in the electromagnetic response of Weyl Hamiltonian~\cite{Vilenkin1980,Nielsen1983}, where the conservation of the chiral charge is violated in presence of both electric and magnetic field~\cite{Nielsen1983}. Interestingly, it was also pointed out that a similar phenomenon can be captured by a semiclassical transport theory, by taking into account the Berry phase effect~\cite{Son2013}. While these studies on the magneto-transport phenomena revealed the effect of quantum anomalies on transport phenomena, experimental investigation in high-energy physics remains unexplored due to the lack of experimentally accessible Weyl fermions.

Theoretical predictions of the Weyl fermions in ferromagnetic metals \cite{Fang2003} and also semimetals, i.e., Weyl semimetals, (WSMs)~\cite{Herring1937,Murakami2007,Burkov2011,Xu2013,Wan2011} opened the possible studies of chiral anomaly in solids. This theoretical possibility is indeed realized by the discovery of three-dimensional Dirac~\cite{Liu2014,Neupane2014} and Weyl~\cite{Lv2015,Xu2015} semimetals. These materials are considered to be an experimental platform for studying unique responses in Dirac and Weyl fermions to the electromagnetic field~\cite{Oka2009,Potter2014,Moll2016,Taguchi2016,Ebihara2016,Chan2016,Ishizuka2016,Wu2016,Ishizuka2017,Chan2017,Ma2017,deJuan2017,Osterhoudt2017,Zhang2018a,Zhang2018b}. Furthermore, the realization in solids allowed experimental generation of pseudo-electromagnetic fields by magnetic fluctuations~\cite{Liu2013} and by lattice strains~\cite{Guinea2009,Levy2010,Chernodub2014,Cortijo2015,Pikulin2016,Sumiyoshi2016,Grushin2016,Cortijo2016,Gorbar2017,Tchoumakov2017,Kariyado2017}, providing greater freedom in experiments to study rich physics related to Weyl semimetals, such as the quantum anomaly.

\begin{figure}
  \centering
  \includegraphics[width=\linewidth]{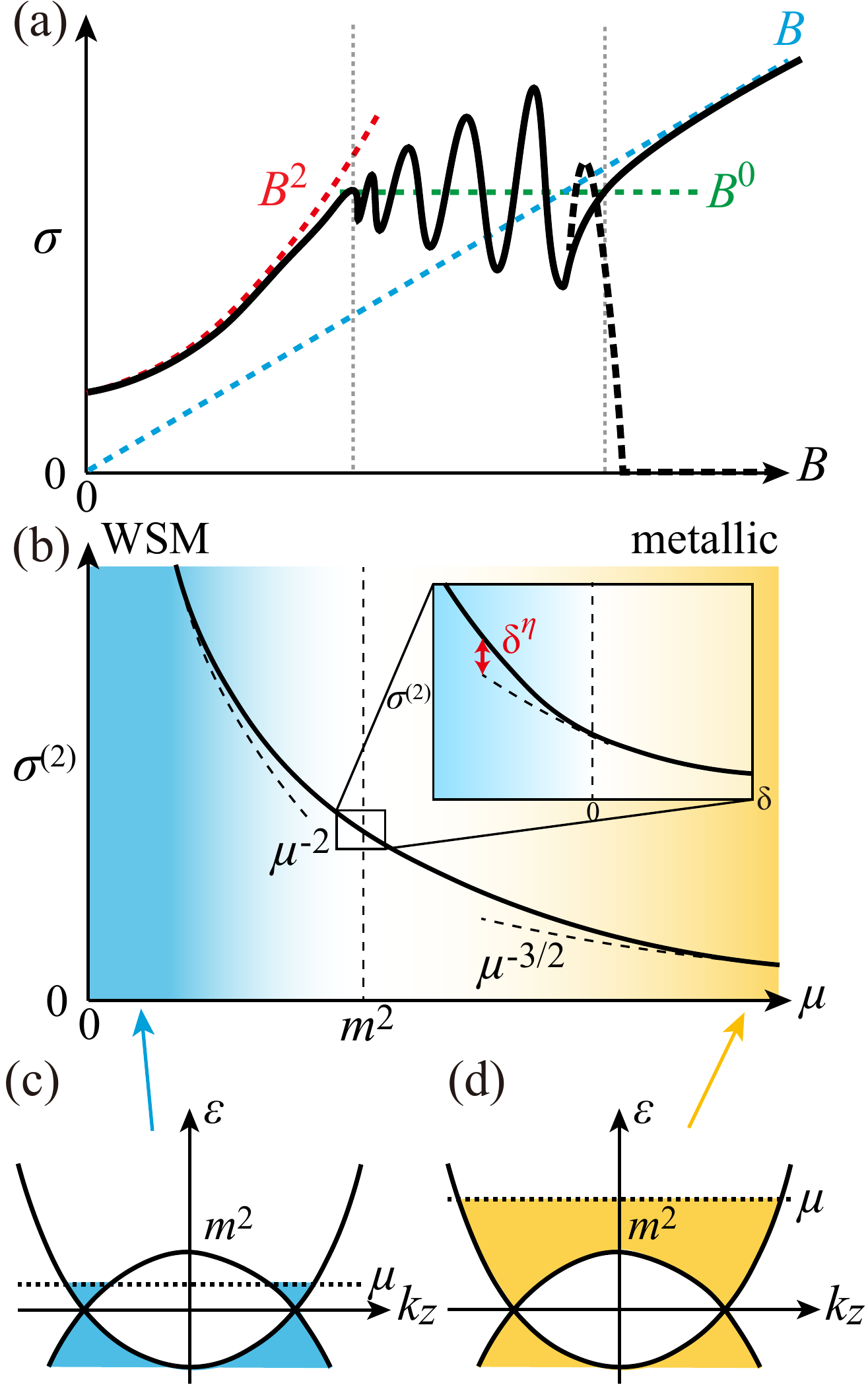}
\caption{The conductivity of Weyl semimetals. (a) Field dependence of the longitudinal conductivity 
when the magnetic field is applied parallel to the electric field. The conductivity shows three regions: 
the semiclassical region under the weak field, quantum oscillation region in the intermediate field, and the 
high-field limit which either the lowest Landau level (solid line) or no Landau level (dashed line) crosses 
the Fermi level [See Fig.~\ref{fig:landau}(b) and related arguments in the main text.]. The behavior of conductivity is for the case in which the chemical potential is much 
higher than the saddle point [the case shown in (d)]. (b) Schematic figure of the chemical potential 
($\mu$) dependence of the ${\cal O}(B^2)$ part of the conductivity in the semiclassical region when 
the electric and magnetic fields are parallel. The chemical potential dependence of the ${\cal O}(B^2)$
 conductivity changes from $\mu^{-2}$ in the $\mu\ll m^2$ to $\mu^{-3/2}$ in the $\mu\gg m^2$ limit. When the fields are along $a=x,y$, it shows non-analytic structure below the saddle point with $\eta=3/2$; such behavior vanishes for $a=z$ when $\bm B\parallel\bm E$. A schematic figure of the two Weyl nodes and the electron filling for $\mu\ll m^2$ and $\mu\gg m^2$ are shown in (c) and (d), respectively.}\label{fig:mBdep}
\end{figure}

In experiments, the chiral anomaly is often investigated by transport experiments where the anomaly is predicted to gives rise to negative longitudinal magnetoresistance (LMR)~\cite{Nielsen1983,Fukushima2008,Son2013,Sekine2017}; the LMR experiment is carried out in several Weyl and Dirac semimetals~\cite{Liang2014,Huang2015,Xiong2015,Li2016b,Zhang2016,Hirschberger2016,Kuroda2017,Niemann2017,Zhang2017}, showing negative magnetoresistance which is seemingly consistent with the theoretical predictions. The experimental confirmation of chiral anomaly, however, still remains a controversial issue. In part, this is a technical problem related to the distinction between different mechanisms for negative LMR, which was recently investigated experimentally~\cite{Liang2018}. A more fundamental problem, on the other hand, remains on the validity of the Weyl Hamiltonian. Unlike its counterpart in the high-energy theory, the realistic effective theory for the existing Weyl and Dirac semimetals turns out to be somewhat complicated than the Weyl Hamiltonian; the Weyl Hamiltonian only applies to a limited energy range often below the Fermi energy. For example, in Cd$_3$As$_2$, the Fermi level of the material is above the saddle point that separates the two nodes [Fig.~\ref{fig:mBdep}(d)]. This situation in the materials cast doubt on how match of the physics of Weyl fermions survives in these materials. Nevertheless, a negative LMR similar to that in other WSM materials is recently observed in Cd$_3$As$_2$~\cite{Li2016,Nishihaya2018}; naively, this implies one of the two possibilities: the physics related to chiral anomaly is robust (appears in a rather general class of models with Weyl nodes) or the observed LMR is not related to the chiral anomaly. Despite the controversial situation, systematic theoretical investigation on such issues remains unexplored.

In this work, we study the general features of anomaly-related magnetoresistance. In particular, we study the anomaly-related LMR when the chemical potential is away from the Weyl nodes, considering both the semiclassical region under a weak magnetic field and quantum region in the presence of strong magnetic field where the Landau levels are well formed. In the semiclassical limit, we derive a general formula for the anomaly-related ${\cal O}(B^2E)$ response, which is related to the Berry curvature. The formula explicitly shows that the contribution to the anomaly-related LMR does not cancel out even when all nodes are enclosed in a Fermi surface, i.e., when the Fermi level is far away from the Weyl nodes. The formula also implies that no abrupt decrease of the anomaly-related current occurs at the Lifshitz point where the Fermi surfaces of the Weyl nodes merge. We demonstrate these results by applying the theory to a model with two Weyl nodes [Fig.~\ref{fig:mBdep}(c) and \ref{fig:mBdep}(d)]. We discuss that the anomaly-related LMR robustly remains even when the Fermi level is far above the saddle point separating the two nodes [Fig.~\ref{fig:mBdep}(d)], and no abrupt decrease occurs at the $\mu=m^2$, when the chemical potential reaches the saddle point of dispersion that connects the two Weyl nodes. This result indicates that the anomaly-related current of similar magnitude should be observed even when the Fermi level is above the saddle point, as long as the chemical potential is in the same order as the energy of the saddle point. On the other hand, the LMR under the strong-magnetic-field case is more dependent on the details of the model; a clear LMR appears only in the strong-field limit, if the lowest Landau level crosses the Fermi level. However, whether the chiral mode (lowest Landau level) crosses the Fermi level depends on the details of the Hamiltonian and the chemical potential. The behavior of the conductivity when the Fermi level is above the saddle point is shown in Fig.~\ref{fig:mBdep}(a), where the black solid line is for the case in which the chiral mode exists at the Fermi level and the dashed line is for the case when the chiral mode is away from the Fermi level. In addition, we find that a remnant LMR exists even when the Weyl nodes vanish by the pair annihilation. We also reveal the critical behavior around this gap opening transition. The result indicates that the negative LMR in the semiclassical region robustly remains, especially in the weak-field limit, providing a strong evidence for the existence of Weyl nodes and/or in close vicinity of the Weyl semimetal (WSM) phase.

\section{Negative magnetoresistance in semiclassical transport theory}\label{sec:LMR}

To investigate the behavior of anomaly-related phenomena when the Fermi level is away from the Weyl nodes, we first consider the case of the weak magnetic-field limit $\omega_c\tau\ll1$, where $\omega_c$ is the cyclotron frequency of the electrons and $\tau$ is the relaxation time. We reformulate the semiclassical Boltzmann theory for magnetoresistance and explicitly show that, if the anomaly-related current exists, the sign of anomaly-related magnetoresistance is negative for arbitrary Hamiltonian. Moreover, the formula we introduce directly shows that the contribution from different nodes do not cancel out even when the Fermi level is away from the Weyl nodes and all nodes are enclosed in a Fermi surface.

We start from the semiclassical Boltzmann theory with Berry phase collection~\cite{Sundaram1999,Xiao2010}. In the Boltzmann theory, the electron density $n_{\bm k\alpha}$ for band $\alpha$ and momentum $\bf k$ is calculated from the Boltzmann equation
\begin{align}
  &\partial_t n_{\bm k\alpha}+ \left(1+q\bm B\cdot\bm b_{\bm k\alpha}\right)^{-1}\nonumber\\
  &\times\left(q\bm E+q\bm v_{\bm k\alpha}\times\bm B+q^2(\bm E\cdot\bm B)\bm b_{\bm k\alpha}\right)\cdot\nabla_{\bm k}n_{\bm k\alpha} = -\frac{\delta n_{\bm k\alpha}}\tau,\label{eq:dn}
\end{align}
where $q$ is the charge of the particle and $\bm E\equiv(E_x,E_y,E_z)$ [$\bm B\equiv(B_x,B_y,B_z)$] is the electric (magnetic) field; $n_{\bm k \alpha}$, $\bm v_{\bm k\alpha}$, and $\bm b_{\bm k\alpha}$ are respectively the occupation, group velocity, and the Berry curvature. We here used the relaxation time approximation where $\tau$ is the relaxation time, $\delta n_{\bm k\alpha}\equiv n_{\bm k\alpha}-n_{\bm k\alpha}^0$ and $n_{\bm k\alpha}^0 = f_0(\varepsilon_{\bm k\alpha})$ is the Fermi-Dirac distribution function with $\varepsilon_{\bm k\alpha}$ being the energy of the state.

To study the general features of the anomaly-related transport phenomena, we generalize the Berry phase formalism developed in Ref.~\onlinecite{Son2013}. Several forms of the generalizations of Ref.~\onlinecite{Son2013} was attempted in several recent works for Weyl Hamiltonians\cite{Kim2014,Lundgren2014,Sharma2016,McCormick2016,Gao2017,Sekine2017,Sharma2017,Wei2018}. However, here, we reformulate the formula in a form which explicitly shows that the LMR from the anomaly-related current always gives a negative contribution to LMR. In the semiclassical Boltzmann theory, the electric current reads
\begin{align}
  \bm J= \sum_\alpha\int \frac{d\bm k}{(2\pi)^3}\left[\bm v_{\bm k\alpha}+q\bm E\times\bm b_{\bm k\alpha}+\frac{q}{c}(\bm b_{\bm k\alpha}\cdot\bm v_{\bm k\alpha})\bm B\right] n_{\bm k \alpha},\label{eq:current}
\end{align}
where $q$ is the charge of the particle and $\bm E\equiv(E_x,E_y,E_z)$ [$\bm B\equiv(B_x,B_y,B_z)$] is the electric (magnetic) field; $n_{\bm k \alpha}$, $\bm v_{\bm k\alpha}$, and $\bm b_{\bm k\alpha}$ are respectively the occupation, group velocity, and the Berry curvature of electrons with crystal momentum $\bm k$ and band index (including spin) $\alpha$. Note that the first term in the integrand of Eq.~\eqref{eq:current} corresponds to the conventional current, the second term to the intrinsic anomalous Hall effect (AHE), while the third term is the anomaly related contribution which is the main interest of the present paper. Assuming the steady state ($\partial_t n_{\bm k\alpha}=0$), and solving the Boltzmann equation in Eq.~\eqref{eq:dn}, we find
\begin{align}
  \delta n_{\bm k\alpha}&= -\tau\left(1+q\bm B\cdot\bm b_{\bm k\alpha}\right)^{-1}\times\nonumber\\
  &\left(q\bm E+q\bm v_{\bm k\alpha}\times\bm B+q^2(\bm E\cdot\bm B)\bm b_{\bm k\alpha}\right)\bm v_{\bm k\alpha}\; (n_{\bm k\alpha}^0)',\label{eq:dn3}
\end{align}
where $(n_{\bm k\alpha}^0)'\equiv \partial_\varepsilon n^0(\varepsilon_{\bm k\alpha})$ is the derivative of the Fermi distribution function. We here expanded the solution up to second order in the electromagnetic fields, and linear order in the relaxation time.

Substituting this equation into Eq.~\eqref{eq:current}, the anomaly-related response in the order of ${\cal O}(EB^2)$ reads
\begin{align}
  \bm J_\text{ano}=&\tau q^4\sum_\alpha\int\frac{dk^3}{(2\pi)^3} \bm W_{\bm k\alpha}[\bm E\cdot\bm W_{\bm k\alpha}]\delta(\varepsilon_{\bm k\alpha}-\mu),\label{eq:J3b}
\end{align}
where $\bm W_{\bm k\alpha}\equiv \bm b_{\bm k\alpha}\times(\bm v_{\bm k\alpha}\times\bm B)$.  Suppose we apply the electric field along a unit vector $\bm e_E$. Then, the current along $\bm e_E$ reads
\begin{align}
  (J_\text{ano}\cdot\bm e_E)=&\tau q^4E\sum_\alpha\int\frac{dk^3}{(2\pi)^3} (\bm e_E\cdot\bm W_{\bm k\alpha})^2\delta(\varepsilon_{\bm k\alpha}-\mu),\label{eq:J3b2}
\end{align}
where $E=|\bm E|$. Therefore, the Berry phase contribution to LMR exists if the region of $\bm k$ with $\bm W_{\bm k\alpha}\ne 0$ has a finite measure on the Fermi surface. This essentially indicates that a nonzero negative LMR appears when $\bm b_{\bm k\alpha}\ne0$, as there is no reason to be $W_{\bm k\alpha}=0$ on the entire Fermi surface unless $b_{\bm k\alpha}=0$. In WSMs, this also implies that the anomaly-related LMR appears even when all Weyl nodes are enclosed inside one Fermi surface; in this case, the total charge of Weyl nodes are zero, but the Berry curvature induced by the Weyl nodes (and $\bm W_{\bm k\alpha}$) are still nonzero, in general. Another interesting consequence of Eq.~\eqref{eq:J3b2} is that the induced current along the electric field direction always gives a negative contribution to the resistivity, i.e., the anomaly-related LMR is always negative. These general features imply that the anomaly-related LMR in WSMs are also robust, regardless of the details of Hamiltonian. Moreover, the result indicates that the contribution from different Weyl nodes does not cancel out even when the Fermi level is away from the Weyl nodes and all nodes are enclosed in one Fermi surface.

We next consider the case in which $\bm E$ and $\bm B$ are perpendicular to each other. Applying $\bm E\cdot\bm B=0$ to Eq.~\eqref{eq:J3b}, we find
\begin{align}
  (&J_\text{ano}\cdot\bm e_E)=\nonumber\\
&\tau q^4E\sum_\alpha\int\frac{dk^3}{(2\pi)^3} (\bm e_E\cdot\bm v_{\bm k\alpha})^2(\bm B\cdot\bm b_{\bm k\alpha})^2\delta(\varepsilon_{\bm k\alpha}-\mu).
\end{align}
Hence, similar to the case of LMR, negative magnetoresistance due to the Berry curvature also appears when $\bm B\perp \bm E$. Therefore, a negative magnetoresistance in the $\bm E\perp \bm B$ can appear even within the free electron approximation. In real materials, however, this contribution competes with the conventional magnetoresistance which gives a positive contribution~\cite{Gao2017}. Therefore, the sign of the magnetoresistance depends on the details of the band structure. We also note that this term remains finite also for the Weyl Hamiltonian in contrast to Ref.~\onlinecite{Son2013}, in which the anomaly-related current vanish when $\bm E\perp \bm B$. This is difference originates from the different approximation used for the electron distribution; we used the standard relaxation-time approximation while Ref.~\onlinecite{Son2013} considered a limit where intra-node scattering is much faster than the inter-node ones. Further discussion on the sensitivity to the electron distribution is given in Appendix~\ref{sec:Weyl}.

In the last, we note that a similar argument on the amount of charge in each electron/hole pockets in the Brillouin zone shows that the rate of charge pumped between different pockets is determined only by the total charge of magnetic monopoles inside the pocket. This implies that the chiral charge pumping by the chiral anomaly is also robust. Details on this argument are given in Appendix~\ref{sec:anomaly}.

\section{Anomaly-related current in the weak magnetic field}

\subsection{Two-node model}

\begin{figure}[tbp]
  \centering
  \includegraphics[width=\linewidth]{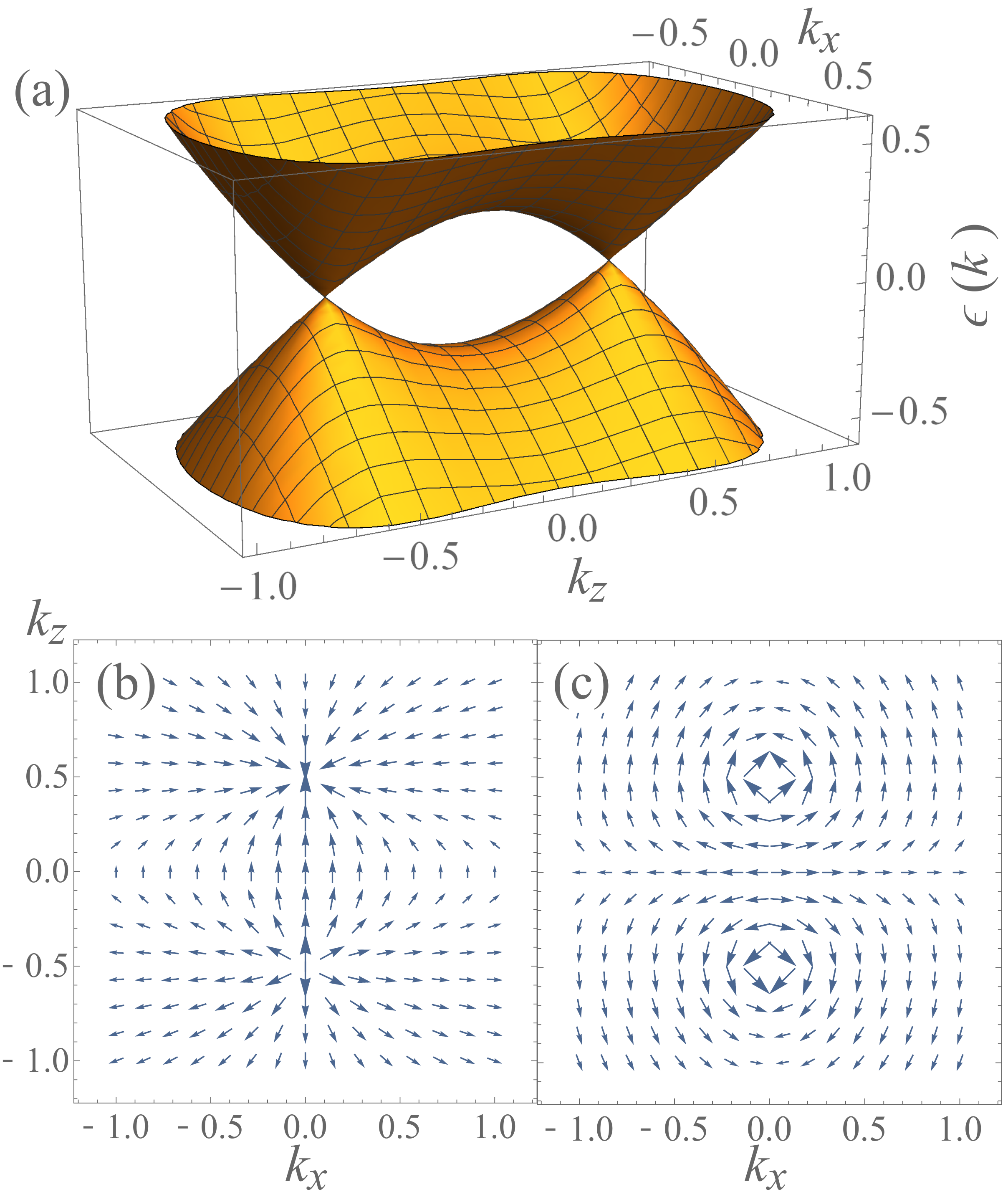}
  \caption{Band structure and Berry curvature distribution in the two-node model. (a) The band structure of the two-node Weyl Hamiltonian. (b) Berry curvature distribution of the two-node Weyl Hamiltonian in the $k_y=0$ plane, and (c) distribution of $\bm W_+(\bm k)$ of the same Hamiltonian. The results are for $m=1/2$.}\label{fig:dipolarWeyl}
\end{figure}

In this section, we apply the general argument presented in Sec.~\ref{sec:LMR} to a model with two Weyl nodes and study how the anomaly-related response typically behaves with changing $\mu$. The Hamiltonian reads
\begin{align}
  H_{\rm DW}=&\int\frac{dk^3}{(2\pi)^3}\psi^\dagger(k)\left\{\sigma_1k^1+\sigma_2k^2\right.\nonumber\\
  &\qquad\qquad\left.+\sigma_3((k^3)^2-m^2)-\mu\sigma_0\right\}\psi(k).\label{eq:Hdw}
\end{align}
When $m^2>0$, this model has two Weyl nodes at $\varepsilon=0$; this is an example of the Weyl semimetals with broken time-reversal symmetry. The low energy theory around the Weyl points become the isotropic Weyl Hamiltonian when $m=1/2$, while the cones become anisotropic otherwise. The dispersion is shown in Fig.~\ref{fig:dipolarWeyl}(a). This model has two Weyl nodes, which are separated by a saddle point; when $|\mu|<m^2,$ there are two Fermi surfaces, each enclosing one Weyl node. On the other hand, the two Fermi surfaces merge into one when $|\mu|>m^2$. The Berry curvature of the conduction band $\bm b_{\bm k+}$ is shown in Fig.~\ref{fig:dipolarWeyl}(b); the source and the drain of $\bm b_{\bm k+}$ at $k_z=\pm1/2$ corresponds to the position of the Weyl nodes. In this model, the Berry curvature distributes in the momentum space in a similar manner as the magnetic field around a magnetic dipole, i.e., it decays in a power of $k=|\bm k|$ when $k\to\infty$. Reflecting this feature of $\bm b_{\bm k+}$, $\bm W_{\bm k+}$ also decays in the power of $k$; the distribution of $\bm W_{\bm k+}$ when the magnetic field is along the $z$ axis is shown in Fig.~\ref{fig:dipolarWeyl}(c). The result shows that $\bm W_{\bm k+}$ remains finite even when $\varepsilon_{\bm k+}>m^2$. According to Eq.~\eqref{eq:J3b2}, this implies that the magnetoresistance related to the Berry curvature also appears even when the Fermi level is $\mu>m^2$. 

The anomaly-related current in some limits of this model was previously studied using a formula similar to Eq.~\eqref{eq:J3b}~\cite{Cortijo2016b}. On the other hand, we calculate the anomaly related ${\cal O}(EB^2)$ response for an arbitrary value of $\mu$ using Eq.~\eqref{eq:J3b}. We find the general solution has the form
\begin{subequations}
\begin{align}
  J_\text{ano}^a=\frac{2\tau q^4}{(2\pi)^3}&\left[\sigma_1^a(\bm B\cdot\bm E)B_a+\sigma_2^aQ_z(\bm E,\bm B)B_a\right.\nonumber\\
&\left.\quad+\sigma_3^a(\bm B\cdot\bm B)E_a+\sigma_4^aQ_z(\bm B,\bm B)E_a\right],\label{eq:J3bxy}
\end{align}
for $a=x,y$ and
\begin{align}
  J_\text{ano}^z=\frac{2\tau q^4}{(2\pi)^3}&\left[\sigma_1^z(\bm B\cdot\bm E)B_z+\sigma_2^zQ_z(\bm E,\bm B)B_z\right.\nonumber\\
&\left.\qquad+\sigma_3^z(\bm B\cdot\bm B)E_z\right],\label{eq:J3bz}
\end{align}\label{eq:J3b3}
\end{subequations}
where
\begin{align}
  Q_z(\bm A,\bm B)=\frac1{\sqrt3}\left(3A_zB_z-\bm A\cdot\bm B\right).
\end{align}
For Eq.~\ref{eq:J3bz}, the term that corresponds to the fourth term in Eq.~\ref{eq:J3bxy} can be converted to a sum of the other three terms, i.e., $Q_z(\bm B,\bm B)E_z=\frac1{\sqrt3}(\bm B\cdot\bm E)B_z+Q_z(\bm E,\bm B)B_z-\frac1{\sqrt3}(\bm B\cdot\bm B)E^z$; therefore, the fourth term is absent in Eq.~\ref{eq:J3bz}. In the previous works considering Weyl Hamiltonian, the current is proportional to $(\bm B\cdot\bm E)\bm B$. In contrast, our result for the anomaly-related current has four terms: $(\bm B\cdot\bm E)\bm B$, $Q_z(\bm B,\bm E)\bm B$, $(\bm B\cdot\bm B)\bm E$, $Q_z(\bm B,\bm B)\bm E$. The general form of $\sigma_i^a$ for arbitrary $\mu$ are given in Appendix~\ref{sec:solution}.

In the $\mu\to0$ limit, the result corresponds to a Weyl semimetal with two Weyl nodes. The solution for this limit is obtained by Laurent expansion of $\sigma_i^a$ by $\mu$ around $\mu=0$ point. The result reads
\begin{subequations}
\begin{align}
  J_\text{ano}^a&=\frac{2\tau q^4}{(2\pi)^3}\left[\frac{14\pi m}{15\mu^2}(\bm B\cdot\bm E)B_a+\frac{\pi(1+8m^2)}{90m\mu^2}(\bm B\cdot\bm B)E_a\right.\nonumber\\
&\qquad\qquad\left.+\frac{\pi(1-4m^2)}{30\sqrt3m\mu^2}Q_z(\bm B,\bm B)E_a\right],
\end{align}
for $a=x,y$ and
\begin{align}
  J_\text{ano}^z&=\frac{2\tau q^4}{(2\pi)^3}\left[\frac{4\pi m(11-2m^2)}{45\mu^2}(\bm B\cdot\bm E)B_z\right.\nonumber\\
&\quad\left.+\frac{2\pi m(1-4m^2)}{15\sqrt3m\mu^2}Q_z(\bm E,\bm B)B_z+\frac{8\pi m^3}{15\mu^2}(\bm B\cdot\bm B)E_z\right].
\end{align}
\end{subequations}
Hence, by approaching the Weyl node ($\mu=0$), the current diverges with $\propto1/\mu^2$. The anisotropy in the above solution reflects the anisotropy of the velocity of the Weyl nodes. In the result, the ratio of $(\bm E\cdot\bm B)\bm B$ and $(\bm B\cdot\bm B)\bm E$ terms reads $\chi^a\equiv\sigma_1^a/\sigma_3^a = 84m^2/(1+8m^2)$ for $a=x,y$ and $\chi^z\equiv\sigma_1^z/\sigma_3^z = (11-2m^2)/(6m^2)$. Hence, the $(\bm E\cdot\bm B)\bm B$ term is either similar or larger than the $(\bm B\cdot\bm B)\bm E$ term when $m$ is the order of 1. The anisotropy is natural since the distribution of Berry curvature is dipole-like and anisotropic in momentum space.

In the Hamiltonian in Eq.~\eqref{eq:Hdw}, the velocity of Weyl nodes is isotropic when $m=1/2$. In this case, the above solutions become isotropic:
\begin{align}
  \bm J_\text{ano}&=\frac{\tau q^4}{4\pi^2c^2}\left[\frac{7}{15\mu^2}(\bm B\cdot\bm E)\bm B+\frac{1}{15\mu^2}(\bm B\cdot\bm B)\bm E\right].\label{eq:J3smallm}
\end{align}
This qualitatively reproduces the results for the Weyl Hamiltonian studied in Ref.~\cite{Son2013}. The result, however, has several differences: we find the current proportional to $(\bm B\cdot\bm B)\bm E$ in addition to the $(\bm B\cdot\bm E)\bm B$ term, and the current is reduced by $8/15$ when $\bm E$ and $\bm B$ are parallel (the ratio of the two terms is $\chi^a = 7$). As discussed in Appendix~\ref{sec:Weyl}, these differences are consequences of the difference in the electron distribution assumed. The difference in the result shows that, unlike the usual Berry phase effects, the anomaly-related current is sensitive to the electron distribution. This is a natural behavior considering that Eq.~\eqref{eq:J3b} indicates the current is related to the change of the electron distribution on the Fermi surface. 

We next turn to the case when $\mu>m^2$, i.e., when the chemical potential is above the saddle point. As mentioned above, as the magnitude of the Berry curvature decays with a power of $\mu$, it is expected that the anomaly-related current also decays with a power of $\mu$, i.e., they remain finite even above the saddle point. Indeed, by expanding the general solution with respect to $1/\mu$, we find the current in the $\mu\gg m^2$ limit reads:
\begin{subequations}
\begin{align}
  J_\text{ano}^a=&\frac{2\tau q^4}{(2\pi)^2}\left[\frac{116}{693\mu^{3/2}}(\bm B\cdot\bm E)B_a-\frac{1}{33\sqrt3\mu^{3/2}}Q_z(\bm E,\bm B)B_a\right.\nonumber\\
      &\left.+\frac{8}{693\mu^{3/2}}(\bm B\cdot\bm B)E_a-\frac{4}{231\sqrt3\mu^{3/2}}Q_z(\bm B,\bm B)E_a\right],
\end{align}
for $a=x,y$ and
\begin{align}
  J_\text{ano}^z=&\frac{2\tau q^4}{(2\pi)^2}\left[-\frac{4(154\mu-975-34m^2)}{27027\mu^{3/2}}(\bm B\cdot\bm E)B_z\right.\nonumber\\
&-\frac{616\mu+78-136m^2}{9009\sqrt3\mu^{3/2}}Q_z(\bm E,\bm B)B_z\nonumber\\
&\left.+\frac{616\mu-136m^2}{9009\mu^{3/2}}(\bm B\cdot\bm B)E_z\right].
\end{align}\label{eq:J3_highm}
\end{subequations}
Therefore, the anomaly related magnetoresistance remains nonzero even when $\mu\gg m^2$. The result, however, shows a qualitative difference in the asymptotic behavior compared to the $\mu\ll m^2$ case. In this limit, the asymptotic form of the current is $J_\text{ano}^a\propto\mu^{-3/2}$ for $a=x,y$ and $J_\text{ano}^z\propto\mu^{-1/2}$, slower than the $J_\text{ano}^a\propto \mu^{-2}$ decay in the $\mu\ll m^2$ case. Therefore the decay of magnetoresistance is slower when the Fermi level is away, compared to the $\mu\ll m^2$ case. One more point to be noted is that, although $J_\text{ano}^z$ generally decays in $\mu^{-1/2}$, the $\mu^{-1/2}$ terms vanish when both $\bm E$ and $\bm B$ fields are applied along $z$ axis, and the leading order becomes $J_\text{ano}^z\propto\mu^{-3/2}B_z^2E_z$. In the $\mu\gg m^2$ limit, $\chi^a=29/2$ for $a=x,y$ and $\chi^z=1/3+325/(34m^2-154\mu)\sim1/3-325/(154\mu)$. Hence, the $(\bm E\cdot\bm B)\bm B$ term is either similar or an order of magnitude larger than the $(\bm B\cdot\bm B)\bm E$ term.

The above argument on the $\mu\gg m^2$ case indicates that the LMR of a similar order to the case of $\mu< m^2$ remains if the current does not decay rapidly when $\mu$ crosses $m^2$, i.e., when the Fermi level crosses the saddle point. We study this possibility by expanding the general solution with $\delta\equiv\mu-m^2$; we find the current has two terms, $\bm J_\text{ano}=\bm J_{b1}+\bm J_{b2}\Theta(-\delta)$, where $\bm J_{b1}$ is the analytic part and the later is the singular part of the current ($\Theta(x)$ is the Heaviside's step function): 
\begin{subequations}
\begin{align}
  J_{b2}^a=&\frac{\tau q^4}{(2\pi)^2m^3}\left[-\frac{4}{9}\left|\frac{\delta}{m^2}\right|^{\frac32}(\bm B\cdot\bm E)B_a\right.\nonumber\\
  &+\frac{2}{3\sqrt3}\left|\frac{\delta}{m^2}\right|^{\frac32}Q_z(\bm E,\bm B)B_a\nonumber\\
  &+\frac1{9m^2}\left|\frac{\delta}{m^2}\right|^{\frac32}(\bm B\cdot\bm B)E_a\nonumber\\
  &\left.+\frac1{3\sqrt3m^2}\left|\frac{\delta}{m^2}\right|^{\frac32}Q_z(\bm B,\bm B)E_a\right],
\end{align}
for $a=x,y$ and
\begin{align}
  J_{b2}^z=&\frac{2\tau q^4}{(2\pi)^2}\left[\frac{4}{15m^3}\left|\frac\delta{m^2}\right|^{\frac52}(\bm B\cdot\bm E)B_z\right.\nonumber\\
  &\qquad-\frac{2}{5\sqrt3m^3}\left|\frac\delta{m^2}\right|^{\frac52}Q_z(\bm E,\bm B)B_z\nonumber\\
  &\left.\qquad+\frac{8}{35m}\left|\frac\delta{m^2}\right|^{\frac72}(\bm B\cdot\bm B)E_z\right].
\end{align}
\end{subequations}
The explicit form of $\bm J_{b1}$ is given in Appendix~\ref{sec:solution}. Reflecting the singular change of the Fermi surface at $\mu=m^2$, $\bm J_\text{ano}$ shows a non-analytic behavior which is characterized by the divergence of the second derivative of $J_\text{ano}^a$ with respect to $\mu$ for $a=x,y$, and third derivative for $J_\text{ano}^z$. However, the result is continuous and changes smoothly when $\mu\sim m^2$. As no abrupt decrease in the magnetoresistance appears around $\mu\sim m^2$, we expect to see the LMR of similar magnitude even when $\mu>m^2$, i.e., when the two Weyl nodes are enclosed in a Fermi surface.

The summary of the above analyses is shown in Fig.~\ref{fig:mBdep}(b) for the case when $\bm B\parallel\bm E$. As the distribution of $\bm W_{\bm k\alpha}$ in Fig.~\ref{fig:dipolarWeyl}(c) is smooth throughout the Brillouin zone, we expect a gradual change of the anomaly-related MR. Indeed, we find the Berry phase related current of this model diverges proportional to $\mu^{-2}$ in the $\mu\to 0$ limit and decays $\mu^{-3/2}$ when $\mu\gg m^2$. These two limits are connected smoothly without any abrupt change at $\mu=m^2$, when the Fermi surfaces of the two Weyl nodes merge. Our result on the two-node Hamiltonian shows that the anomaly-related MR robustly remains even when the Fermi level is far away from the Weyl nodes.

\subsection{Anomaly-related current near the critical point}
\begin{figure}
  \centering
  \includegraphics[width=\linewidth]{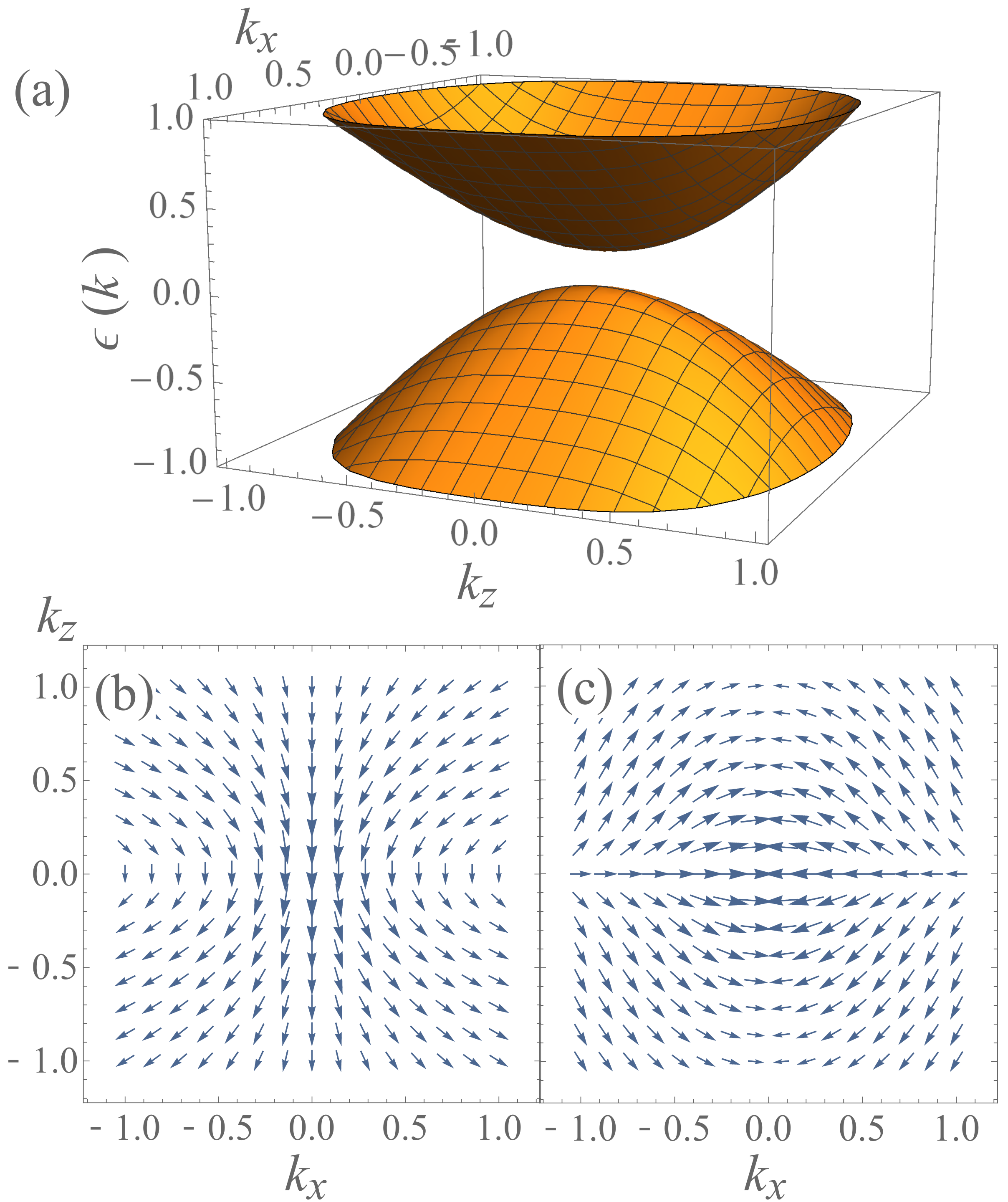}
  \caption{Band structure and Berry curvature distribution in the gapped Hamiltonian $H_g$. (a) The band structure, (b) Berry curvature distribution in the $k_y=0$ plane, and (c) distribution of $\bm W_+(\bm k)$. The results are for $m=1/3$.}\label{fig:gapped}
\end{figure}

We next investigate how the anomaly-related current behaves around the critical point at which the two Weyl nodes pair annihilates. As an example of such, we here consider the model similar to Eq.~\eqref{eq:Hdw}, but with opposite sign for the $m^2$ term (the two Weyl nodes in Eq.~\eqref{eq:Hdw} merge at $m^2=0$):
\begin{align}
  H_{\rm g}=&\int\frac{dk^3}{(2\pi)^3}\psi^\dagger(k)\left\{\sigma_1k^1+\sigma_2k^2\right.\nonumber\\
  &\qquad\qquad\left.+\sigma_3((k^3)^2+m^2)-\mu\sigma_0\right\}\psi(k).\label{eq:Hdw2}
\end{align}
Here, we assumed $m^2>0$. The band structure of the model is shown in Fig.~\ref{fig:gapped}(a). This model shows a band gap of size $2m^2$; the density of states is zero when $\mu\in[-m^2,m^2]$. Although the band structure looks like a trivial semiconductor, the conduction band of this model has non-zero Berry curvature as shown in Fig.~\ref{fig:gapped}(b), and hence, non-zero $\bm W_{\bm k+}$ as in Fig.~\ref{fig:gapped}(c). Therefore, although no Weyl nodes exist, we expect a similar LMR even after the Weyl nodes pair annihilates, at least close to the pair annihilation point where the Berry curvature around the band bottom is large.

When $m^2=0$, i.e., at the critical point~\cite{Yang2014}, the anomaly-related current in the weak magnetic-field limit is obtained by substituting $m^2=0$ to Eqs.~\eqref{eq:sigma_gapped} and \eqref{eq:sigma_gappedz}:
\begin{subequations}
\begin{align}
	J_\text{ano}^a=&\frac{\tau q^4}{(2\pi)^2}\left\{\frac{232}{693\mu^{\frac32}}(\bm B\cdot\bm E)B_a-\frac{2}{33\sqrt3\mu^{\frac32}}Q_z(\bm E,\bm B)B_a\right.\nonumber\\
  &\left.+\frac{77+240\mu}{10395\mu^{\frac52}}(\bm B\cdot\bm B)E_a+\frac{77-120\mu}{3465\sqrt3\mu^{\frac52}}Q_z(\bm B,\bm B)E_a\right\},
\end{align}
for $a=x,y$ and
\begin{align}
	J_\text{ano}^z=&\frac{\tau q^4}{(2\pi)^2}\left\{-\frac{2(154\mu-975)}{27027\mu^{\frac32}}(\bm B\cdot\bm E)B_a\right.\nonumber\\
  &\left.-\frac{39+308\mu}{9009\sqrt3\mu^{\frac32}}Q_z(\bm E,\bm B)B_a+\frac{4}{117\mu^{\frac12}}(\bm B\cdot\bm B)E_a\right\}.
\end{align}
\end{subequations}
We note that, when $\bm B\parallel\bm E$, only the current along the electric field direction survives with chemical potential dependence $\mu^{-3/2}$. Hence, when $\mu\to0$, the current diverges with a different power from the Weyl case, $\propto\mu^{-2}$.

When $m^2>0$ and $\mu$ is close to the band edge, i.e., $\mu\sim m^2$, the current reads:
\begin{subequations}
\begin{align}
	J_\text{ano}^a=&\frac{\tau q^4}{(2\pi)^2}\left[\frac{4}{9m^6}\delta^{\frac32}(\bm B\cdot\bm E)B_a-\frac{2}{3\sqrt3m^6}\delta^{\frac32}Q_z(\bm E,\bm B)B_a\right.\nonumber\\
        &\left.+\frac{1}{9m^8}\delta^{\frac32}(\bm B\cdot\bm B)E_a+\frac{1}{3\sqrt3m^8}\delta^{\frac32}Q_z(\bm B,\bm B)E_a\right],
\end{align}
for $a=x,y$ and
\begin{align}
	J_\text{ano}^z=&\frac{\tau q^4}{5\pi^2m^8}\left[\frac{2}{3}\delta^{\frac52}(\bm B\cdot\bm E)B_z-\frac{1}{\sqrt3}\delta^{\frac52}Q_z(\bm E,\bm B)B_z\right.\nonumber\\
  &\left.+\frac{4}{7}\delta^{\frac72}(\bm B\cdot\bm B)E_z\right],\label{eq:Jb3gapped}
\end{align}
\end{subequations}
for $\delta\equiv \mu-m^2\ge0$ and zero otherwise. Therefore, a finite anomaly-related current appears when $\delta>0$. The critical behavior of the current is highly anisotropic reflecting the symmetry of the Hamiltonian; it increases by $\delta^{\frac32}$ for the current along $x$ and $y$. For the $z$ axis, in general, the current is proportional to $\delta^{\frac52}$ for the current along $z$ axis. However, when $\bm B$ and $\bm E$ are both parallel to the $z$ axis, the first two terms in Eq.~\eqref{eq:Jb3gapped} cancels, and the leading order becomes $\delta^{\frac72}$. When $\delta<0$, on the other hand, the Fermi level is in the band gap, and therefore, the current is zero.

The result for $\bm J_\text{ano}$ for $\mu\gg m^2$ is the same as in Eq.~\eqref{eq:J3_highm} for $J_\text{ano}^x$ and $J_\text{ano}^y$. For $J_\text{ano}^z$, the term proportional to $m^2$ differs from the Weyl semimetal case:
\begin{align}
	J_\text{ano}^z=\frac{2\tau q^4}{(2\pi)^2}&\left[-\frac{4(154\mu-975-351m^2)}{27027\mu^{3/2}}(\bm B\cdot\bm E)B_z\right.\nonumber\\
&\qquad-\frac{616\mu+78-1404m^2}{9009\sqrt3\mu^{3/2}}Q_z(\bm E,\bm B)B_z\nonumber\\
&\qquad\left.+\frac{616\mu-1404m^2}{9009\mu^{3/2}}(\bm B\cdot\bm B)E_z\right].
\end{align}
Hence, the Berry-curvature-related current exists for arbitrary filling, even after the Weyl nodes vanish. The remnant of the Weyl nodes is found in the magnetoresistance even after the two Weyl nodes vanish by a pair annihilation.

\begin{figure}
  \centering
  \includegraphics[width=\linewidth]{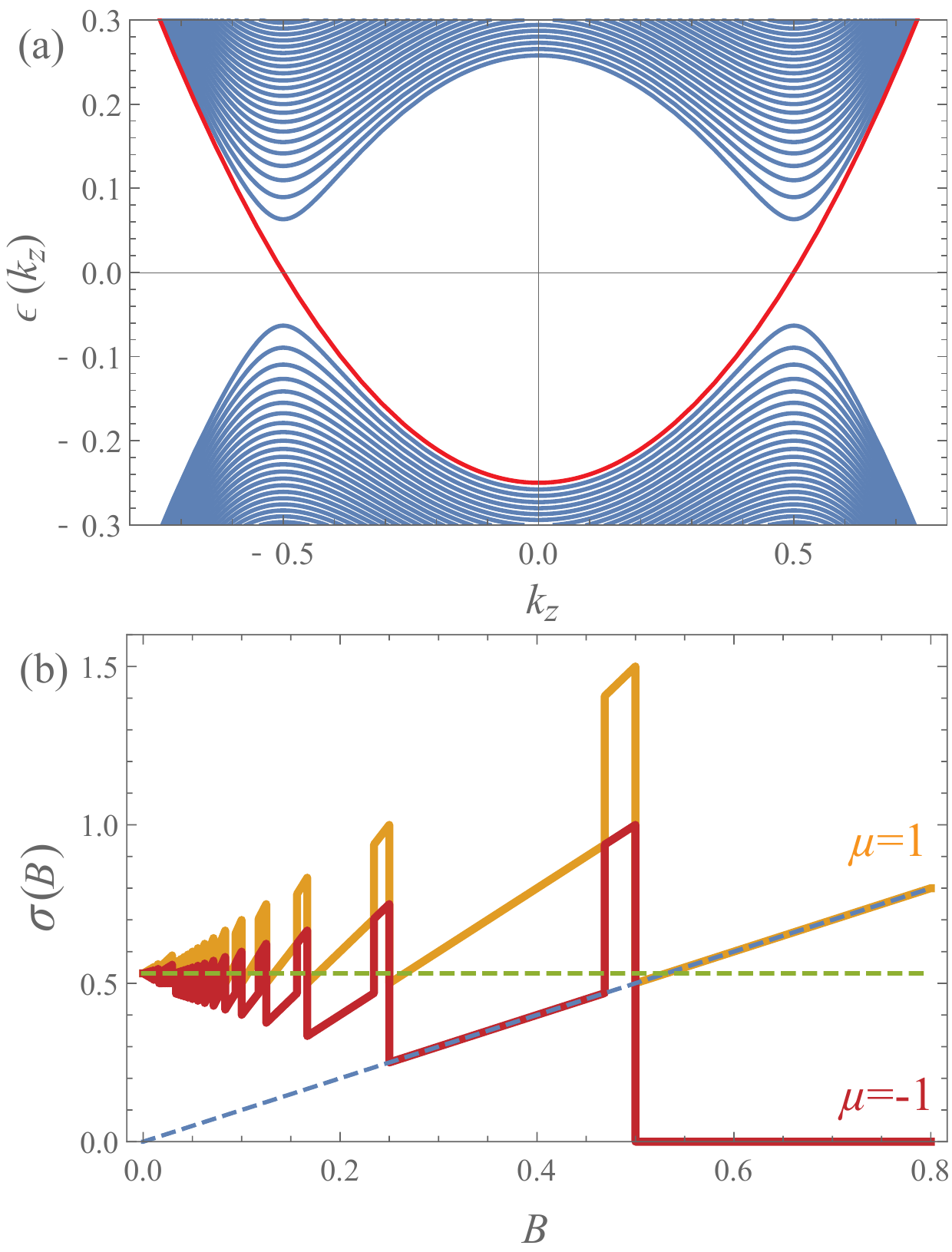}
  \caption{Electronic and transport properties under the strong magnetic field. (a) Landau levels of the Hamiltonian in Eq.~\eqref{eq:Hdw}. The result is for $m=1/2$ and when the magnetic field is applied along the $z$ axis. The Weyl nodes are located at $k_z=\pm1/2$ when no magnetic field is applied. The red band corresponds to the chiral mode of Weyl Hamiltonian. (b) Magnetic field ($B$) dependence of electric current along the $z$ axis for $m=1/2$ and $\mu=1$ (yellow) and $-1$ (red). The green dashed lines indicates the conductivity in the limit $\sqrt{B}\ll\mu$ and the blue dashed line corresponds to the case in which only the chiral mode crosses the Fermi level. When $\mu=-1$, the conductivity goes to zero in the strong $B$ limit, as the Fermi level is below the bottom of the chiral mode.}\label{fig:landau}
\end{figure}

\section{Anomaly-related current under the strong field}
\subsection{Two-node model}
We next turn to the strong field limit $\omega_c\tau\gg1$, where the Landau levels are formed. In this limit, the above semiclassical treatment of the magnetic field fails. Instead, we study the MR in this limit by considering the Boltzmann theory for the Landau levels~\cite{Nielsen1983}. We here consider the case in which the magnetic field is applied along the $z$ axis. In this case, the eigenenergies and the eigenstate wavefunctions of the Landau levels for the general Hamiltonian
\begin{align}
  H=&\int\frac{dk^3}{(2\pi)^3}\psi^\dagger(k)\left\{\sigma_1k^1+\sigma_2k^2+\sigma_3f(k_z)-\mu\sigma_0\right\}\psi(k).\label{eq:H2}
\end{align}
is obtained by the same method used for Weyl Hamiltonian in Ref.~\cite{Nielsen1983}; here, $f(k_z)$ is a general function of $k_z$. Assuming $qB_z>0$, the eigenenergy reads
\begin{align}
  \omega_n={\rm sgn}(n)\sqrt{2qB_z|n|+f^2(k_z)},\label{eq:omegaLL}
\end{align}
where ${\rm sgn}(n)=1$ for $n\ge0$ and $-1$ for $n<0$; most of the results remain the same when $qB<0$, except that ${\rm sgn}(n)=-1$ for $n=0$. Here, each Landau levels are $N_\omega=\lfloor \frac{qB_z}{2\pi}L^2\rfloor$ fold degenerate, where $L$ is the length along $x$ and $y$ directions and $\lfloor x\rfloor$ is the largest integer smaller than $x$. Therefore, when a Landau level crosses the Fermi level, the contribution of the level to the conductivity is expected to increase proportionally to $B$.

The Landau levels for the model in Eq.~\ref{eq:Hdw}, i.e., $f(k_z)=k_z^2-m^2$, is shown in Fig.~\ref{fig:landau}. The mode that crosses the $\epsilon=0$ corresponds to the chiral modes of the Weyl Hamiltonian; $\omega_0(k_z)=0$ when $k_z=\pm m$, which correspond to the positions of the Weyl nodes in absence of the magnetic field. When $\mu^2<2qB_z$, the chiral mode is the only mode that contributes to the electric conduction. In this limit, the Boltzmann theory within the relaxation-time approximation gives $j_z = \frac{q^3}{2\pi^2}\tau E_z B_z,$ corresponds to the two Weyl nodes case in Ref.~\cite{Nielsen1983}; the result is independent of the group velocity of the electrons, which is a characteristic feature of the 1d systems.

When $\mu^2>2qB_z$, the result for $\mu^2<2qB_z$ is modified to include the contributions from different Landau levels:
\begin{align}
j_z = \frac{q^3n_b}{2\pi^2}\tau E_z B_z,
\end{align}
where $2n_b$ is the number of crossings between the Landau levels and the Fermi level with
\begin{align}
  n_b=& \left[2\left\lfloor\frac{\mu^2}{2qB_z}\right\rfloor - \left\lfloor\frac{\mu^2-m^2}{2qB_z}\right\rfloor\Theta(\mu^2-m^2)+\Theta(\mu+m)\right];\label{eq:nlandau}
\end{align}
$\lfloor x\rfloor$ is the floor function. The last term in the square bracket comes from the chiral mode. The quantum oscillation comes from the product of $B_z$ and the floor functions in $n_b$. The result is plotted in Fig.~\ref{fig:landau}(b). In the $\mu^2\gg m^2, 2qB_z$ limit, Eq.~\eqref{eq:nlandau} is approximated as $n_b=(\mu^2+m^2)/(2qB_z)+{\cal O}(1)$, where the ${\cal O}(1)$ is related to the quantum oscillation. Therefore, the current becomes
\begin{align}
j_z \sim \frac{q^2}{4\pi^2}\tau (\mu^2+m^2)E_z. \label{eq:J_highm2}
\end{align}
Hence, in this limit, the magnetic field dependence of conductivity only shows up in the oscillation, which is the ${\cal O}(1)$ effect neglected in Eq.~\eqref{eq:J_highm2}. In Fig.~\ref{fig:landau}(b), this corresponds to the limit $B\to0$, where the conductance converges to a finite value with oscillations around it. A notable difference in the structure of the quantum oscillation appears in the box-like structure of the conductivity at the right-end of the slope. This reflects the band structure of the two-node model, where the bottom of each Landau level is located at $k_z\ne0$; this structure vanishes when $\mu\le m^2$.

Another characteristic behavior appears in the strong field limit $\sqrt{qB}>|\mu|$. In this limit, the Landau levels with $|n|>0$ does not cross the Fermi level as the modes with $n>0$ ($n<0$) move to a higher (lower) energy. Therefore, the only Landau level that can cross the Fermi level is the $n=0$ mode. This is the case when $\mu>-m^2$ of which an example is shown in Fig.~\eqref{fig:landau}(b) for $\mu=1$; this corresponds to the situation considered in Ref.~\cite{Nielsen1983,Fukushima2008}. For $\mu<-m^2$, however, the system shows a different behavior. In the two-node model in Eq.~\eqref{eq:Hdw}, the chiral mode exists only for $\mu\ge-m^2$. Therefore, when the Fermi level is $\mu<m^2$, the conductivity goes to zero in the strong field limit. These results are summarized in Fig.~\ref{fig:mBdep}(a).  

When the magnetic field is flipped, i.e., $qB<0$, the dispersion of the chiral mode also flips, i.e. $\omega_0=-(k_z^2-m^2)$, and therefore, the equation for $n_b$ become
\begin{align}
  n_b=& \left[2\left\lfloor\frac{\mu^2}{2qB_z}\right\rfloor - \left\lfloor\frac{\mu^2-m^2}{2qB_z}\right\rfloor\Theta(\mu^2-m^2)-\Theta(\mu+m)\right].\label{eq:nlandau2}
\end{align}
Hence, in this case, the negative magnetoresistance in the strong-field limit appears only when $\mu<m^2$, in the opposite to the $qB>0$ cases.

\subsection{Near critical point}

A similar argument also holds for the model in Eq.~\eqref{eq:Hdw2}, when the system is near the critical point which the two Weyl nodes appear from a band touching. In this case, the current reads:
\begin{align}
j_z =&\frac{q^3}{2\pi^2}\tau E_z B_z\left[\left\lfloor\frac{\mu^2-m^4}{2qB_z}\right\rfloor\Theta(\mu^2-m^4)+\Theta(\mu-m^2)\right].
\end{align}
Most of the arguments above also apply to this case: when $\mu\gg m,qB$, the conductivity is $\sigma_{zz}=\frac{q^2\tau}{4\pi^2}(\mu^2-m^4)+{\cal O}(1)$ where the quantum oscillation is included in the ${\cal O}(1)$, and $\sigma_{zz}\propto B_z$ in the strong field limit when only the lowest Landau level crosses the Fermi level. A difference is that the chiral mode crosses the Fermi level only when $\mu>m^2$ for $qB>0$ ($\mu<-m^2$ for $qB<0$), in contrast to $\mu>-m^2$ ($\mu<m^2$) in the case with two Weyl nodes.

\section{Discussion}
To conclude, we studied the magnetoresistance in the Weyl semimetals that are related to the chiral anomaly. In the semiclassical limit, using the Boltzmann theory approach, we derive a general formula for the ${\cal O}(B^2E)$ response in the weak-field limit which is related to the Berry curvature of the electronic bands. The result shows the anomaly-related current is proportional to the square of $W_{\bm k\alpha}^a$, as shown in Eq.~\eqref{eq:J3b2}; for longitudinal negative magnetoresistance, it always gives an additional current along the electric field direction, i.e., longitudinal negative magnetoresistance as $(W_{\bm k\alpha}^a)^2\ge0$. In the case of Weyl semimetals, this shows that the contribution from different Weyl nodes do not cancel each other even when the Fermi level is away from the Weyl nodes and all the nodes are enclosed in a single Fermi surface. By considering a model with two Weyl nodes, we explicitly show that the anomaly-related magnetoresistance exists even when multiple Weyl nodes share a Fermi surface. Moreover, we find that the anomaly-related MR decays smoothly with increasing chemical potential. Therefore, a similar magnitude of magnetoresistance is expected even when all nodes are in a pocket.

On the other hand, in the quantum limit where the Landau levels are well formed, the behavior of the conductivity is governed by the structure and the degeneracy of Landau levels. In the relatively weak field region where a large number of Landau levels cross the Fermi surface, we show that the conductivity is roughly independent of the magnetic field unlike the semiclassical case; the quantum oscillation appears as a small wiggle on top of the field-independent conductivity. The chiral magnetic effect studied in Ref.~\cite{Fukushima2008} is evident in the conductivity only when the lowest Landau level (chiral mode) crosses the Fermi surface. In general, this depends on the details of the Hamiltonian and chemical potential. However, the chiral mode in Eq.~\eqref{eq:Hdw} crosses the Fermi level for arbitrary $\mu>-m^2$ (if $qB>0$). Therefore, the contribution from the chiral mode may also appear even if the Fermi level is far away from the Weyl nodes.

These results show that, especially in the semiclassical region, the magnetoresistance related to the chiral anomaly is extremely robust regardless of the details of the Hamiltonian nor the position of the Fermi surface. Furthermore, in our analyses on the gapped model in Eq.~\eqref{eq:Hdw2} (see also Fig.~\eqref{fig:gapped}), we find that the magnetoresistance remains even after the Weyl nodes pair annihilates.

In regard to experiments, recent experiments on Dirac/Weyl semimetal candidates observe negative magnetoresistance. However, its relation to the anomaly-related magnetoresistance is still under debate. One of the concerns in these experiments was the location of the Fermi level; it is often far away from the Weyl nodes and a Fermi surface encloses multiple Weyl nodes. Our theory, however, shows that the magnetoresistance should remain finite even in such cases. This is consistent with the recent experiments in Cd$_3$As$_2$~\cite{Li2016,Nishihaya2018}.

Another point regarding the experiment is the existence of multiple pairs of Weyl nodes. The unconventional electromagnetic response related to Weyl fermions often cancels out when taking into account of the existence of multiple nodes. For the chiral anomaly studied here, in the semiclassical limit, the total contribution from all pairs of Weyl fermions can be calculated simply by summing the contribution from each pair. Furthermore, as discussed in Eq.~\eqref{eq:J3b2}, each pair always gives a negative contribution to the resistivity. Therefore, no cancellation takes place. For instance, in Cd$_3$As$_2$~\cite{Wang2013,Neupane2014}, there are two Dirac nodes, i.e., two pairs of Weyl nodes which are degenerate. The total current induced by the chiral anomaly in this material is given by the sum of contributions from the two pairs.

A slightly complicated example of Weyl semimetal is the pyrochlore iridates, where a Weyl semimetal phase is expected in the vicinity of the phase boundary of metal-insulator transition~\cite{Wan2011,Witczak-Krempa2013}, which is realized by controlling the temperature~\cite{Matsuhira2011}, by applying magnetic field~\cite{Ueda2015,Tian2015}, or by chemical substitution~\cite{Ueda2015b}. In this material, there are four pairs of Weyl nodes of which a pair resides on each $\langle111\rangle$ direction; they are related by the cubic symmetry of the magnetically ordered phase. In this material, when the Fermi level is sufficiently above the Weyl nodes, all eight nodes are expected to be in the same Fermi surface. This is a situation where the above results on the two-node model do not directly apply. Furthermore, the anti-ferroic pattern of the position of the eight Weyl nodes generates a multipolar pattern of the Berry curvature in the momentum space, which may seem to cancel the anomaly-related phenomena. However, from the general argument on Eq.~\eqref{eq:J3b2}, we see that the cancellation does not occur as the anomaly-related current is proportional to the momentum integral of $|\bm W_{\bm k\alpha}|^2$ where $\bm W_{\bm k\alpha}$ is defined below Eq.~\eqref{eq:J3b}.

\acknowledgements

The authors thank M. Kawasaki, J. Liu, Y. Nakazawa, S. Nishihaya, and M. Uchida for fruitful discussions. We particularly thank M. Kawasaki and M. Uchida for bringing our attention to this problem. This work was supported by JSPS KAKENHI (Grant Nos. JP16H06717, JP18H03676, JP18H04222, and JP26103006), ImPACT Program of Council for Science, Technology and Innovation (Cabinet Office, Government of Japan), and CREST, JST (Grant No. JPMJCR16F1).

\appendix

\section{Pumping of charge between different electron/hole pockets by chiral anomaly}\label{sec:anomaly}
We here show that, in the semiclassical limit, the rate of charges pumped by the chiral anomaly is determined only by the total charge of Weyl nodes enclosed in the pocket, regardless of the Hamiltonian nor the shape of the pocket. We consider the case $\tau\to\infty$. In this case, the change of $n_{\bm k\alpha}$ become
\begin{align}
  &\partial_t n_{\bm k\alpha}=- \left(1+q\bm B\cdot\bm b_{\bm k\alpha}\right)^{-1}\nonumber\\
  &\times\left(q\bm E+q\bm v_{\bm k\alpha}\times\bm B+q^2(\bm E\cdot\bm B)\bm b_{\bm k\alpha}\right)\cdot\nabla_{\bm k}n_{\bm k\alpha}.\label{eq:dn2}
\end{align}
Intuitively, the $q^2(\bm E\cdot\bm B)\bm b_{\bm k\alpha}$ term in Eq.~\eqref{eq:dn2} represents the change of wavepacket momentum; it flows along the Berry curvature $\bm b_{\bm k\alpha}$. In the case of the Weyl Hamiltonian, this gives a current flowing outward from/inward to the Weyl node, as the Weyl nodes are source/sink of $\bm b_{\bm k\alpha}$; this violates the chiral-charge conservation, which is known as the chiral anomaly~\cite{Nielsen1983,Son2013}. In general Hamiltonian, the time derivative of electron number in the $i$th Fermi surface reads
\begin{align}
  \dot N^{(i)}_\alpha \equiv& \int_{BZ^{(i)}} \frac{dk^3}{(2\pi)^3}\left(1+q\bm B\cdot\bm b_{\bm k\alpha}\right)\partial_tn_{\bm k\alpha},\\
  =& \int_{\partial V^{(i)}} \frac{dS}{(2\pi)^3}\bm n_S\cdot\left(q\bm v_{\bm k\alpha}\times\bm B+q^2(\bm E\cdot\bm B)\bm b_{\bm k\alpha}\right),\\
  =& \frac{q^2}{(2\pi)^3}(\bm E\cdot\bm B)C^{(i)}.
\end{align}
Here, $BZ^{(i)}$ is a part of the Brillouin zone that encloses the region inside the $i$th Fermi surface $V^{(i)}\subset BZ^{(i)}$, $\partial V^{(i)}$ is the Fermi surface of $V^{(i)}$, $\bm n_S$ is the unit vector perpendicular to the surface $dS$. From the first to the second equation, we used the Gauss law. In the second equation, the first term in the integrand is zero as $\bm v_{\bm k\alpha}\parallel \bm n_S$. Hence, $\dot N^{(i)}_\alpha$ is proportional to the total charge of Weyl nodes $C^{(i)}\equiv\int_{\partial V^{(i)}} dS\,(\bm n_S\cdot\bm b_{\bm k\alpha})$, i.e., the amount of charge pumped between different Fermi surfaces is related to the topological property of the Weyl nodes. Therefore, the above argument on the ``breakdown'' of the chiral-charge conservation generally holds for arbitrary Hamiltonian whenever $C^{(i)}$ is nonzero. This implies that the chiral anomaly remains robust in Weyl semimetals, as the key feature of the Weyl Hamiltonian that gives rise to the chiral anomaly is the topological charge defined by $\bm b_{\bm k\alpha}$; the chiral anomaly occurs whenever the nonzero $C^{(i)}$ exist inside the Fermi surface, regardless of the precise form of Hamiltonian, location of the Weyl nodes, nor how far the chemical potentials are from them.

\section{Sensitiveness of the magnetoresistance to the electron distribution}\label{sec:Weyl}

We demonstrate how the details of electron distribution affect the anomaly-related current. For this purpose, we consider the Weyl Hamiltonian
\begin{align}
  H_{W}=v\bm\sigma\cdot\bm p,
\end{align}
where the sign of $v$ corresponds to the different chiralities of the Weyl electrons. For the electron distribution, we consider:
\begin{align}
  \delta n_{\bm k\alpha}=&-q\tau\left\{1-q\bm B\cdot\bm b_{\bm p}+\left(q\bm B\cdot\bm b_{\bm p}\right)^2\right\}(\bm E\cdot\bm v_{\bm p})\; (n_{\bm p}^0)'\nonumber\\
  &-q^2\left\{\tau_{int}-q\tau\bm B\cdot\bm b_{\bm p}\right\}(\bm E\cdot\bm B)(\bm b_{\bm p}\cdot\bm v_{\bm p})\; (n_{\bm p}^0)'.
\end{align}
This is a slightly generalized version of the expansion of $\delta n_{\bm k\alpha}$ in Eq.~\eqref{eq:dn3}; $\tau$ for one of the term is replaced by $\tau_{int}$; $\tau_{int}$ corresponds to the inter-node scattering considered in Ref.~\onlinecite{Son2013}. When $\tau_{int}=\tau$, the electron distribution corresponds to the relaxation time approximation considered in this work, and $\tau=0$ corresponds to the approximation employed in Ref.~\cite{Son2013}; in the later case, the electron distribution remains symmetric while it becomes asymmetric in the former case due to the electron acceleration by the external field. The two relaxation times are phenomenologically introduced to show how the result changes when different approximation for the relaxation time (or electron distribution) is used.

Using these results, the ${\cal O}(EB^2)$ current reads
\begin{align}
	\bm J_\text{ano} = \frac{q^3|v|^3}{8\pi^2\mu^2}\left\{\left(\tau_{int}-\frac8{15}\tau\right)(\bm E\cdot\bm B)\bm B + \frac\tau{15}(\bm B\cdot\bm B)\bm E\right\}.
\end{align}
When $\tau=0$, the current is proportional to $(\bm E\cdot\bm B)\bm B$, reproducing the result in Ref.~\cite{Son2013}. When $\tau>0$, the result shows another term proportional to $(\bm B\cdot\bm B)\bm E$, which comes from the asymmetry of the electron distribution around the Fermi surface.

\section{Solution of anomaly-related current in two-node Hamiltonian}\label{sec:solution}
\subsection{Solution of anomaly-related current for general $\mu$}
We here present the general solution for the anomaly related nonlinear current in Eq.~\eqref{eq:J3b} for the Hamiltonian in Eq.~\eqref{eq:Hdw}. The calculation was performed by reducing the three-dimensional integral in Eq.~\eqref{eq:J3b} to the integral over the Fermi surface. For arbitrary $\mu$, we find the coefficients $\sigma_i^a$ reads (see Eq.~\eqref{eq:J3b3} for the definition of $\sigma_i^a$):
\begin{widetext}
\begin{subequations}
\begin{align}
\sigma_1^a&= \left\{
\begin{array}{ll}
\frac{4\pi(m^2-|\mu|)^{\frac12}}{10395}\left(\frac{32m^{10}}{|\mu|^7}+\frac{16m^{8}}{|\mu|^6}-\frac{186m^6}{|\mu|^5}-\frac{89m^4}{|\mu|^4}-\frac{643m^2}{|\mu|^3}+\frac{870}{|\mu|^2}\right)\\
\qquad+\frac{4\pi(m^2+|\mu|)^{\frac12}}{10395}\left(-\frac{32m^{10}}{|\mu|^7}+\frac{16m^{8}}{|\mu|^6}+\frac{186m^6}{|\mu|^5}-\frac{89m^4}{|\mu|^4}+\frac{643m^2}{|\mu|^3}+\frac{870}{|\mu|^2}\right) & (|\mu|\le m^2)\\
\frac{4\pi(m^2+|\mu|)^{\frac32}}{10395}\left(-\frac{32m^{8}}{|\mu|^7}+\frac{48m^6}{|\mu|^6}+\frac{138m^4}{|\mu|^5}-\frac{227m^{2}}{|\mu|^4}+\frac{870}{|\mu|^3}\right)& (|\mu|> m^2)
\end{array}\right.\\
\sigma_2^a&= \left\{
\begin{array}{ll}
\frac{2\pi(m^2-|\mu|)^{\frac12}}{693\sqrt3}\left(\frac{32m^{10}}{|\mu|^7}+\frac{16m^{8}}{|\mu|^6}+\frac{12m^6}{|\mu|^5}+\frac{10m^4}{|\mu|^4}-\frac{49m^2}{|\mu|^3}-\frac{21}{|\mu|^2}\right)\\
\qquad+\frac{2\pi(m^2+|\mu|)^{\frac12}}{693\sqrt3}\left(-\frac{32m^{10}}{|\mu|^7}+\frac{16m^{8}}{|\mu|^6}-\frac{12m^6}{|\mu|^5}+\frac{10m^4}{|\mu|^4}+\frac{49m^2}{|\mu|^3}-\frac{21}{|\mu|^2}\right) & (|\mu|\le m^2)\\
-\frac{2\pi}{693\sqrt3}(m^2+|\mu|)^{\frac32}\left(\frac{32m^{8}}{|\mu|^7}-\frac{48m^6}{|\mu|^6}+\frac{60m^4}{|\mu|^5}-\frac{70m^{2}}{|\mu|^4}+\frac{21}{|\mu|^3}\right)& (|\mu|> m^2)
\end{array}\right.\\
\sigma_3^a&= \left\{
\begin{array}{ll}
\frac{\pi(m^2-|\mu|)^{\frac12}}{10395}\left(-\frac{64m^{10}-352m^8}{|\mu|^7}-\frac{32m^{8}-176m^6}{|\mu|^6}+\frac{240m^6-330m^4}{|\mu|^5}+\frac{112m^4-121m^2}{|\mu|^4}-\frac{496m^2+77}{|\mu|^3}+\frac{240}{|\mu|^2}\right)\\
\qquad+\frac{\pi(m^2+|\mu|)^{\frac12}}{10395}\left(\frac{64m^{10}-352m^8}{|\mu|^7}-\frac{32m^{8}-176m^6}{|\mu|^6}-\frac{240m^6-330m^4}{|\mu|^5}+\frac{112m^4-121m^2}{|\mu|^4}+\frac{496m^2+77}{|\mu|^3}+\frac{240}{|\mu|^2}\right) & (|\mu|\le m^2)\\
\frac{\pi}{10395}(m^2+|\mu|)^{\frac32}\left(\frac{64m^{8}-352m^6}{|\mu|^7}-\frac{96m^6-528m^4}{|\mu|^6}-\frac{144m^4+198m^2}{|\mu|^5}+\frac{256m^{2}+77}{|\mu|^4}+\frac{240}{|\mu|^3}\right)& (|\mu|> m^2)
\end{array}\right.
\end{align}
\begin{align}
\sigma_4^a&= \left\{
\begin{array}{ll}
\frac{\pi(m^2-|\mu|)^{\frac12}}{3465\sqrt3}\left(\frac{32m^{10}+352m^8}{|\mu|^7}+\frac{16m^{8}+176m^6}{|\mu|^6}-\frac{120m^6+330m^4}{|\mu|^5}-\frac{56m^4+121m^2}{|\mu|^4}+\frac{248m^2-77}{|\mu|^3}-\frac{120}{|\mu|^2}\right)\\
\qquad+\frac{\pi(m^2+|\mu|)^{\frac12}}{3465\sqrt3}\left(-\frac{32m^{10}+352m^8}{|\mu|^7}+\frac{16m^{8}+176m^6}{|\mu|^6}+\frac{120m^6+330m^4}{|\mu|^5}-\frac{56m^4+121m^2}{|\mu|^4}-\frac{248m^2-77}{|\mu|^3}-\frac{120}{|\mu|^2}\right) & (|\mu|\le m^2)\\
-\frac{\pi}{3465\sqrt3}(m^2+|\mu|)^{\frac32}\left(\frac{32m^{8}+352m^6}{|\mu|^7}-\frac{48m^6+528m^4}{|\mu|^6}-\frac{72m^4-198m^2}{|\mu|^5}+\frac{128m^{2}-77}{|\mu|^4}+\frac{120}{|\mu|^3}\right)& (|\mu|> m^2)
\end{array}\right.
\end{align}
\end{subequations}
for $a=x, y$ and
\begin{subequations}
\begin{align}
\sigma_1^z&= \left\{
\begin{array}{ll}
\frac{8\pi}{135135}(m^2-|\mu|)^{\frac12}\left(-\frac{192m^{12}-416m^{10}}{|\mu|^7}-\frac{96m^{10}-208m^8}{|\mu|^6}+\frac{500m^8+1014m^6}{|\mu|^5}+\frac{226m^6+559m^4}{|\mu|^4}\right.\\
\left.\qquad\qquad+\frac{162m^4-7072m^2}{|\mu|^3}-\frac{1370m^2-4875}{|\mu|^2}+\frac{770}{|\mu|}\right)\\
\qquad+\frac{8\pi}{135135}(m^2+|\mu|)^{\frac12}\left(\frac{192m^{12}-416m^{10}}{|\mu|^7}-\frac{96m^{10}-208m^8}{|\mu|^6}-\frac{500m^8+1014m^6}{|\mu|^5}+\frac{226m^6+559m^4}{|\mu|^4}\right.\\
\left.\qquad\qquad-\frac{162m^4-7072m^2}{|\mu|^3}-\frac{1370m^2-4875}{|\mu|^2}-\frac{770}{|\mu|}\right)& (|\mu|\le m^2)\\
\frac{8\pi(m^2+|\mu|)^{\frac52}}{135135}\left(\frac{192m^{8}-416m^6}{|\mu|^7}-\frac{480m^6-1040m^4}{|\mu|^6}+\frac{268m^4-2678m^2}{|\mu|^5}+\frac{170m^{2}+4875}{|\mu|^4}-\frac{770}{|\mu|^3}\right)& (|\mu|> m^2)
\end{array}\right.
\end{align}
\begin{align}
\sigma_2^z&= \left\{
\begin{array}{ll}
-\frac{4\pi}{45045\sqrt3}(m^2-|\mu|)^{\frac12}\left(\frac{384m^{12}+1664m^{10}}{|\mu|^7}+\frac{192m^{10}+832m^8}{|\mu|^6}-\frac{1000m^8+3666m^6}{|\mu|^5}-\frac{452m^6+1625m^4}{|\mu|^4}\right.\\
\left.\qquad\qquad-\frac{324m^4-2600m^2}{|\mu|^3}+\frac{2740m^2+195}{|\mu|^2}-\frac{1540}{|\mu|}\right)\\
\qquad-\frac{4\pi}{45045\sqrt3}(m^2+|\mu|)^{\frac12}\left(-\frac{384m^{12}+1664m^{10}}{|\mu|^7}+\frac{192m^{10}+832m^8}{|\mu|^6}+\frac{1000m^8+3666m^6}{|\mu|^5}-\frac{452m^6+1625m^4}{|\mu|^4}\right.\\
\left.\qquad\qquad+\frac{324m^4-2600m^2}{|\mu|^3}+\frac{2740m^2+195}{|\mu|^2}+\frac{1540}{|\mu|}\right) & (|\mu|\le m^2)\\
\frac{4\pi(m^2+|\mu|)^{\frac52}}{45045\sqrt3}\left(\frac{384m^{8}+1664m^6}{|\mu|^7}-\frac{960m^6+4160m^4}{|\mu|^6}+\frac{536m^4+2990m^2}{|\mu|^5}+\frac{340m^{2}-195}{|\mu|^4}-\frac{1540}{|\mu|^3}\right)& (|\mu|> m^2)
\end{array}\right.\\
\sigma_3^z&= \left\{
\begin{array}{ll}
\frac{16\pi}{45045}(m^2-|\mu|)^{\frac12}\left(\frac{96m^{12}}{|\mu|^7}+\frac{48m^{10}}{|\mu|^6}-\frac{250m^8}{|\mu|^5}-\frac{113m^6}{|\mu|^4}-\frac{81m^4}{|\mu|^3}+\frac{685m^2}{|\mu|^2}-\frac{385}{|\mu|}\right)\\
\qquad+\frac{16\pi}{45045}(m^2+|\mu|)^{\frac12}\left(-\frac{96m^{12}}{|\mu|^7}+\frac{48m^{10}}{|\mu|^6}+\frac{250m^8}{|\mu|^5}-\frac{113m^6}{|\mu|^4}+\frac{81m^4}{|\mu|^3}+\frac{685m^2}{|\mu|^2}+\frac{385}{|\mu|}\right) & (|\mu|\le m^2)\\
\frac{16\pi(m^2+|\mu|)^{\frac72}}{45045}\left(-\frac{96m^6}{|\mu|^7}+\frac{336m^4}{|\mu|^6}-\frac{470m^2}{|\mu|^5}+\frac{385}{|\mu|^4}\right)& (|\mu|> m^2)
\end{array}\right.
\end{align}
\end{subequations}
\end{widetext}
The results are the same for $\mu<0$ and $\mu>0$, as $\bm W_{\bm k+}=-\bm W_{\bm k-}$ and the current is proportional to the square of $\bm W_{\bm k\pm}$, where $+$ and $-$ are respectively the conduction and valence bands of the Weyl node. We, however, note that the results of the Boltzmann theory are invalid when the Fermi level is close to the node, i.e., $|\mu|\lesssim q\tau|\bm E|$. This is a consequence of the band crossing, which its effect is not fully taken into account in the Boltzmann theory. The asymptotic form of $J^a_\text{ano}$ shown in the main text is calculated from these analytic solutions.

\subsection{Anomaly-related current around $\mu\sim m^2$}
In the Hamiltonian in Eq.~\eqref{eq:Hdw}, the two Weyl nodes are separated by a saddle point at $\bm k=\bm 0$ and $\mu=m^2$, below which there exist two Fermi surfaces each contains a Weyl node and the two surfaces merge above. We here investigate how the two regions of $\mu$ connect at $\mu=m^2$. An explicit calculation shows $\sigma_i^a$ is a $C_1$ class function at $\mu=m^2$ for $a=x, y$ and $C_2$ class for $\sigma_i^z$ with an exception of $\sigma_3^z$ ($C_3$ class).

By substituting $\mu$ by $\delta\equiv m^2-\mu$ and expanding $\sigma_i^a$, we find
\begin{subequations}
\begin{align}
  \sigma_1^a&=\frac{\sqrt2\pi}{m^3}\left\{\frac{6376}{10395}-\frac{14866}{10395}\frac{\delta}{m^2}\right\}-\frac{4\pi}{9m^3}\left|\frac{\delta}{m^2}\right|^{\frac32}\Theta(-\delta)\\
  \sigma_2^a&=\frac{\sqrt2\pi}{\sqrt3m^3}\left\{\frac{20}{693}+\frac{13}{99}\frac{\delta}{m^2}\right\}+\frac{2\pi}{3\sqrt3m^3}\left|\frac{\delta}{m^2}\right|^{\frac32}\Theta(-\delta)\\
  \sigma_3^a&=\frac{\sqrt2\pi}{m^3}\left\{\frac{22+128m^2}{2079m^2}+\frac{77-2624m^2}{2079m^2}\frac{\delta}{m^2}\right\}\nonumber\\
  &\qquad\qquad+\frac{\pi}{9m^5}\left|\frac{\delta}{m^2}\right|^{\frac32}\Theta(-\delta)\\
  \sigma_4^a&=\frac{\sqrt2\pi}{\sqrt3m^3}\left\{\frac{22-64m^2}{693m^2}+\frac{77+1312m^2}{6930m^2}\frac{\delta}{m^2}\right\}\nonumber\\
  &\qquad\qquad+\frac{\pi}{3\sqrt3m^5}\left|\frac{\delta}{m^2}\right|^{\frac32}\Theta(-\delta)
\end{align}
\end{subequations}
for $a=x, y$ and
\begin{widetext}
\begin{subequations}
\begin{align}
  \sigma_1^z&=-\frac{\sqrt2\pi}{m}\left\{\frac{19840m^2-90272}{135135 m^2}+\frac{189176-33632m^2}{135135m^2}\left(\frac\delta{m^2}\right)+\frac{51588m^2-265499}{135135 m^2}\left(\frac\delta{m^2}\right)^2\right\}+\frac{8\pi}{15m^3}\left|\frac\delta{m^2}\right|^{\frac52}\Theta(-\delta)\\
  \sigma_2^z&=-\frac{\sqrt2\pi}{\sqrt3m}\left\{\frac{39680m^2-9568}{90090m^2}+\frac{15496-67264m^2}{90090m^2}\left(\frac\delta{m^2}\right)+\frac{103176 m^2-38389}{90090m^2}\left(\frac\delta{m^2}\right)^2\right\}-\frac{4\pi}{5\sqrt3m^3}\left|\frac\delta{m^2}\right|^{\frac52}\Theta(-\delta)\\
  \sigma_3^z&=\frac{\sqrt2\pi}m\left\{\frac{3968}{9009}-\frac{33632}{45045}\frac{\delta}{m^2}+\frac{5732}{5005}\left(\frac{\delta}{m^2}\right)^2-\frac{9137}{6435}\left(\frac{\delta}{m^2}\right)^3\right\}+\frac{16\pi}{35m}\left|\frac\delta{m^2}\right|^{\frac72}\Theta(-\delta).
\end{align}
\end{subequations}
\end{widetext}
These results show that a singular behavior appears at $\mu=m^2$, where the derivatives of $\bm J_\text{ano}$ with respect to $\mu$ diverges; the second derivative diverges for $J_\text{ano}^x$ and $J_\text{ano}^y$, while the third derivative diverges for $J_\text{ano}^z$.

\section{Magnetoresistance of the gapped model}
We here present the general solution for the nonlinear current in Eq.~\eqref{eq:J3b} for the Hamiltonian in Eq.~\eqref{eq:Hdw2}. By the same procedure with that of the Weyl semimetal case, we find the coefficients $\sigma_i^a$ are:
\begin{widetext}
\begin{subequations}
\begin{align}
  \sigma_1^a&= \frac{4\pi(\mu-m^2)^{\frac32}}{10395}\left(-\frac{32m^8}{\mu^7}-\frac{48m^6}{\mu^6}+\frac{138m^4}{\mu^5}+\frac{227m^2}{\mu^4}+\frac{870}{\mu^3}\right)\\
\sigma_2^a&=-\frac{2\pi(\mu-m^2)^{\frac32}}{693\sqrt3}\left(\frac{32m^8}{\mu^7}+\frac{48m^6}{\mu^6}+\frac{60m^4}{\mu^5}+\frac{70m^2}{\mu^4}+\frac{21}{\mu^3}\right)\\
\sigma_3^a&= \frac{\pi(\mu-m^2)^{\frac32}}{10395}\left(\frac{64m^8+352m^6}{\mu^7}+\frac{96m^6+528m^4}{\mu^6}-\frac{144m^4-198m^2}{\mu^5}-\frac{256m^2-77}{\mu^4}+\frac{240}{\mu^3}\right)\\
\sigma_4^a&=-\frac{\pi(\mu-m^2)^{\frac32}}{3465\sqrt3}\left(\frac{32m^8-352m^6}{\mu^7}+\frac{48m^6-528m^4}{\mu^6}-\frac{72m^4+198m^2}{\mu^5}-\frac{128m^2+77}{\mu^4}+\frac{120}{\mu^3}\right)
\end{align}\label{eq:sigma_gapped}
\end{subequations}
for $a=x, y$ and
\begin{subequations}
\begin{align}
  \sigma_1^z&=-\frac{8\pi(\mu-m^2)^{\frac52}}{135135}\left(-\frac{192m^8+416m^6}{\mu^7}-\frac{480m^6+1040m^4}{\mu^6}-\frac{268m^4+2678m^2}{\mu^5}+\frac{170m^2-4875}{\mu^4}+\frac{770}{\mu^3}\right),\\
  \sigma_2^z&=-\frac{4\pi(\mu-m^2)^{\frac52}}{45045\sqrt3}\left(-\frac{384m^8-1664m^6}{\mu^7}-\frac{960m^6-4160m^4}{\mu^6}-\frac{536m^4-2990m^2}{\mu^5}+\frac{340m^2+195}{\mu^4}+\frac{1540}{\mu^3}\right),\\
  \sigma_3^z&= \frac{16\pi(\mu-m^2)^{\frac72}}{45045}\left(\frac{96m^6}{\mu^7}+\frac{336m^4}{\mu^6}+\frac{470m^2}{\mu^5}+\frac{385}{\mu^4}\right).
\end{align}\label{eq:sigma_gappedz}
\end{subequations}
\end{widetext}
Further analysis of the above results is presented in the last part of the weak magnetic field section in Results.


\begin{thebibliography}{76}%
\makeatletter
\providecommand \@ifxundefined [1]{%
 \@ifx{#1\undefined}
}%
\providecommand \@ifnum [1]{%
 \ifnum #1\expandafter \@firstoftwo
 \else \expandafter \@secondoftwo
 \fi
}%
\providecommand \@ifx [1]{%
 \ifx #1\expandafter \@firstoftwo
 \else \expandafter \@secondoftwo
 \fi
}%
\providecommand \natexlab [1]{#1}%
\providecommand \enquote  [1]{``#1''}%
\providecommand \bibnamefont  [1]{#1}%
\providecommand \bibfnamefont [1]{#1}%
\providecommand \citenamefont [1]{#1}%
\providecommand \href@noop [0]{\@secondoftwo}%
\providecommand \href [0]{\begingroup \@sanitize@url \@href}%
\providecommand \@href[1]{\@@startlink{#1}\@@href}%
\providecommand \@@href[1]{\endgroup#1\@@endlink}%
\providecommand \@sanitize@url [0]{\catcode `\\12\catcode `\$12\catcode
  `\&12\catcode `\#12\catcode `\^12\catcode `\_12\catcode `\%12\relax}%
\providecommand \@@startlink[1]{}%
\providecommand \@@endlink[0]{}%
\providecommand \url  [0]{\begingroup\@sanitize@url \@url }%
\providecommand \@url [1]{\endgroup\@href {#1}{\urlprefix }}%
\providecommand \urlprefix  [0]{URL }%
\providecommand \Eprint [0]{\href }%
\providecommand \doibase [0]{http://dx.doi.org/}%
\providecommand \selectlanguage [0]{\@gobble}%
\providecommand \bibinfo  [0]{\@secondoftwo}%
\providecommand \bibfield  [0]{\@secondoftwo}%
\providecommand \translation [1]{[#1]}%
\providecommand \BibitemOpen [0]{}%
\providecommand \bibitemStop [0]{}%
\providecommand \bibitemNoStop [0]{.\EOS\space}%
\providecommand \EOS [0]{\spacefactor3000\relax}%
\providecommand \BibitemShut  [1]{\csname bibitem#1\endcsname}%
\let\auto@bib@innerbib\@empty
\bibitem [{\citenamefont {Fujikawa}\ and\ \citenamefont
  {Suzuki}(2004)}]{Fujikawa2004}%
  \BibitemOpen
  \bibfield  {author} {\bibinfo {author} {\bibfnamefont {Kazuo}\ \bibnamefont
  {Fujikawa}}\ and\ \bibinfo {author} {\bibfnamefont {Hiroshi}\ \bibnamefont
  {Suzuki}},\ }\href@noop {} {\emph {\bibinfo {title} {Path integrals and
  quantum anomalies}}}\ (\bibinfo  {publisher} {Oxford Science Publications},\
  \bibinfo {year} {2004})\BibitemShut {NoStop}%
\bibitem [{\citenamefont {Fukuda}\ and\ \citenamefont
  {Miyamoto}(1949)}]{Fukuda1949a}%
  \BibitemOpen
  \bibfield  {author} {\bibinfo {author} {\bibfnamefont {H.}~\bibnamefont
  {Fukuda}}\ and\ \bibinfo {author} {\bibfnamefont {Y.}~\bibnamefont
  {Miyamoto}},\ }\bibfield  {title} {\enquote {\bibinfo {title} {On the
  $\gamma$-decay of neutral meson},}\ }\href@noop {} {\bibfield  {journal}
  {\bibinfo  {journal} {Prog. Theor. Phys.}\ }\textbf {\bibinfo {volume} {4}},\
  \bibinfo {pages} {235} (\bibinfo {year} {1949})}\BibitemShut {NoStop}%
\bibitem [{\citenamefont {Fukuda}\ \emph
  {et~al.}(1949{\natexlab{a}})\citenamefont {Fukuda}, \citenamefont {Miyamoto},
  \citenamefont {Miyajima},\ and\ \citenamefont {Tomonaga}}]{Fukuda1949b}%
  \BibitemOpen
  \bibfield  {author} {\bibinfo {author} {\bibfnamefont {H.}~\bibnamefont
  {Fukuda}}, \bibinfo {author} {\bibfnamefont {Y.}~\bibnamefont {Miyamoto}},
  \bibinfo {author} {\bibfnamefont {T.}~\bibnamefont {Miyajima}}, \ and\
  \bibinfo {author} {\bibfnamefont {S.}~\bibnamefont {Tomonaga}},\ }\bibfield
  {title} {\enquote {\bibinfo {title} {Application of {P}auli's regulator to
  the $\gamma$-decay of neutrettos},}\ }\href@noop {} {\bibfield  {journal}
  {\bibinfo  {journal} {Prog. Theor. Phys.}\ }\textbf {\bibinfo {volume} {4}},\
  \bibinfo {pages} {385} (\bibinfo {year} {1949}{\natexlab{a}})}\BibitemShut
  {NoStop}%
\bibitem [{\citenamefont {Fukuda}\ \emph
  {et~al.}(1949{\natexlab{b}})\citenamefont {Fukuda}, \citenamefont {Miyamoto},
  \citenamefont {Miyazima}, \citenamefont {Tomonaga}, \citenamefont {Oneda},
  \citenamefont {Ozaki},\ and\ \citenamefont {Sasaki}}]{Fukuda1949c}%
  \BibitemOpen
  \bibfield  {author} {\bibinfo {author} {\bibfnamefont {H.}~\bibnamefont
  {Fukuda}}, \bibinfo {author} {\bibfnamefont {Y.}~\bibnamefont {Miyamoto}},
  \bibinfo {author} {\bibfnamefont {T.}~\bibnamefont {Miyazima}}, \bibinfo
  {author} {\bibfnamefont {S.}~\bibnamefont {Tomonaga}}, \bibinfo {author}
  {\bibfnamefont {S.}~\bibnamefont {Oneda}}, \bibinfo {author} {\bibfnamefont
  {S.}~\bibnamefont {Ozaki}}, \ and\ \bibinfo {author} {\bibfnamefont
  {S.}~\bibnamefont {Sasaki}},\ }\bibfield  {title} {\enquote {\bibinfo {title}
  {Applicability of {P}auli's regulator to the $\gamma$-decay of neutrettos},}\
  }\href {\doibase 10.1143/ptp/4.4.477} {\bibfield  {journal} {\bibinfo
  {journal} {Progress of Theoretical Physics}\ }\textbf {\bibinfo {volume}
  {4}},\ \bibinfo {pages} {477--484} (\bibinfo {year}
  {1949}{\natexlab{b}})}\BibitemShut {NoStop}%
\bibitem [{\citenamefont {Steinberger}(1949)}]{Steinberger1949}%
  \BibitemOpen
  \bibfield  {author} {\bibinfo {author} {\bibfnamefont {J.}~\bibnamefont
  {Steinberger}},\ }\bibfield  {title} {\enquote {\bibinfo {title} {On the use
  of subtraction fields and the lifetimes of some types of meson decay},}\
  }\href {\doibase 10.1103/PhysRev.76.1180} {\bibfield  {journal} {\bibinfo
  {journal} {Phys. Rev.}\ }\textbf {\bibinfo {volume} {76}},\ \bibinfo {pages}
  {1180--1186} (\bibinfo {year} {1949})}\BibitemShut {NoStop}%
\bibitem [{\citenamefont {Vilenkin}(1980)}]{Vilenkin1980}%
  \BibitemOpen
  \bibfield  {author} {\bibinfo {author} {\bibfnamefont {Alexander}\
  \bibnamefont {Vilenkin}},\ }\bibfield  {title} {\enquote {\bibinfo {title}
  {Equilibrium parity-violating current in a magnetic field},}\ }\href
  {\doibase 10.1103/PhysRevD.22.3080} {\bibfield  {journal} {\bibinfo
  {journal} {Phys. Rev. D}\ }\textbf {\bibinfo {volume} {22}},\ \bibinfo
  {pages} {3080--3084} (\bibinfo {year} {1980})}\BibitemShut {NoStop}%
\bibitem [{\citenamefont {Nielsen}\ and\ \citenamefont
  {Ninomiya}(1983)}]{Nielsen1983}%
  \BibitemOpen
  \bibfield  {author} {\bibinfo {author} {\bibfnamefont {H.B.}\ \bibnamefont
  {Nielsen}}\ and\ \bibinfo {author} {\bibfnamefont {Masao}\ \bibnamefont
  {Ninomiya}},\ }\bibfield  {title} {\enquote {\bibinfo {title} {The
  {A}dler-{B}ell-{J}ackiw anomaly and {W}eyl fermions in a crystal},}\ }\href
  {\doibase https://doi.org/10.1016/0370-2693(83)91529-0} {\bibfield  {journal}
  {\bibinfo  {journal} {Physics Letters B}\ }\textbf {\bibinfo {volume}
  {130}},\ \bibinfo {pages} {389 -- 396} (\bibinfo {year} {1983})}\BibitemShut
  {NoStop}%
\bibitem [{\citenamefont {Son}\ and\ \citenamefont {Spivak}(2013)}]{Son2013}%
  \BibitemOpen
  \bibfield  {author} {\bibinfo {author} {\bibfnamefont {D.~T.}\ \bibnamefont
  {Son}}\ and\ \bibinfo {author} {\bibfnamefont {B.~Z.}\ \bibnamefont
  {Spivak}},\ }\bibfield  {title} {\enquote {\bibinfo {title} {Chiral anomaly
  and classical negative magnetoresistance of {W}eyl metals},}\ }\href
  {\doibase 10.1103/PhysRevB.88.104412} {\bibfield  {journal} {\bibinfo
  {journal} {Phys. Rev. B}\ }\textbf {\bibinfo {volume} {88}},\ \bibinfo
  {pages} {104412} (\bibinfo {year} {2013})}\BibitemShut {NoStop}%
\bibitem [{\citenamefont {Fang}\ \emph {et~al.}(2003)\citenamefont {Fang},
  \citenamefont {Nagaosa}, \citenamefont {Takahashi}, \citenamefont {Asamitsu},
  \citenamefont {Mathieu}, \citenamefont {Ogasawara}, \citenamefont {Yamada},
  \citenamefont {Kawasaki}, \citenamefont {Tokura},\ and\ \citenamefont
  {Terakura}}]{Fang2003}%
  \BibitemOpen
  \bibfield  {author} {\bibinfo {author} {\bibfnamefont {Zhong}\ \bibnamefont
  {Fang}}, \bibinfo {author} {\bibfnamefont {Naoto}\ \bibnamefont {Nagaosa}},
  \bibinfo {author} {\bibfnamefont {Kei~S.}\ \bibnamefont {Takahashi}},
  \bibinfo {author} {\bibfnamefont {Atsushi}\ \bibnamefont {Asamitsu}},
  \bibinfo {author} {\bibfnamefont {Roland}\ \bibnamefont {Mathieu}}, \bibinfo
  {author} {\bibfnamefont {Takeshi}\ \bibnamefont {Ogasawara}}, \bibinfo
  {author} {\bibfnamefont {Hiroyuki}\ \bibnamefont {Yamada}}, \bibinfo {author}
  {\bibfnamefont {Masashi}\ \bibnamefont {Kawasaki}}, \bibinfo {author}
  {\bibfnamefont {Yoshinori}\ \bibnamefont {Tokura}}, \ and\ \bibinfo {author}
  {\bibfnamefont {Kiyoyuki}\ \bibnamefont {Terakura}},\ }\bibfield  {title}
  {\enquote {\bibinfo {title} {The anomalous {H}all effect and magnetic
  monopoles in momentum space},}\ }\href {\doibase 10.1126/science.1089408}
  {\bibfield  {journal} {\bibinfo  {journal} {Science}\ }\textbf {\bibinfo
  {volume} {302}},\ \bibinfo {pages} {92--95} (\bibinfo {year}
  {2003})}\BibitemShut {NoStop}%
\bibitem [{\citenamefont {Herring}(1937)}]{Herring1937}%
  \BibitemOpen
  \bibfield  {author} {\bibinfo {author} {\bibfnamefont {Conyers}\ \bibnamefont
  {Herring}},\ }\bibfield  {title} {\enquote {\bibinfo {title} {Accidental
  degeneracy in the energy bands of crystals},}\ }\href {\doibase
  10.1103/PhysRev.52.365} {\bibfield  {journal} {\bibinfo  {journal} {Phys.
  Rev.}\ }\textbf {\bibinfo {volume} {52}},\ \bibinfo {pages} {365--373}
  (\bibinfo {year} {1937})}\BibitemShut {NoStop}%
\bibitem [{\citenamefont {Murakami}(2007)}]{Murakami2007}%
  \BibitemOpen
  \bibfield  {author} {\bibinfo {author} {\bibfnamefont {Shuichi}\ \bibnamefont
  {Murakami}},\ }\bibfield  {title} {\enquote {\bibinfo {title} {Phase
  transition between the quantum spin {H}all and insulator phases in 3d:
  emergence of a topological gapless phase},}\ }\href
  {http://stacks.iop.org/1367-2630/9/i=9/a=356} {\bibfield  {journal} {\bibinfo
   {journal} {New Journal of Physics}\ }\textbf {\bibinfo {volume} {9}},\
  \bibinfo {pages} {356} (\bibinfo {year} {2007})}\BibitemShut {NoStop}%
\bibitem [{\citenamefont {Burkov}\ and\ \citenamefont
  {Balents}(2011)}]{Burkov2011}%
  \BibitemOpen
  \bibfield  {author} {\bibinfo {author} {\bibfnamefont {A.~A.}\ \bibnamefont
  {Burkov}}\ and\ \bibinfo {author} {\bibfnamefont {Leon}\ \bibnamefont
  {Balents}},\ }\bibfield  {title} {\enquote {\bibinfo {title} {{W}eyl
  semimetal in a topological insulator multilayer},}\ }\href {\doibase
  10.1103/PhysRevLett.107.127205} {\bibfield  {journal} {\bibinfo  {journal}
  {Phys. Rev. Lett.}\ }\textbf {\bibinfo {volume} {107}},\ \bibinfo {pages}
  {127205} (\bibinfo {year} {2011})}\BibitemShut {NoStop}%
\bibitem [{\citenamefont {Xu}\ \emph {et~al.}(2011)\citenamefont {Xu},
  \citenamefont {Weng}, \citenamefont {Wang}, \citenamefont {Dai},\ and\
  \citenamefont {Fang}}]{Xu2013}%
  \BibitemOpen
  \bibfield  {author} {\bibinfo {author} {\bibfnamefont {Gang}\ \bibnamefont
  {Xu}}, \bibinfo {author} {\bibfnamefont {Hongming}\ \bibnamefont {Weng}},
  \bibinfo {author} {\bibfnamefont {Zhijun}\ \bibnamefont {Wang}}, \bibinfo
  {author} {\bibfnamefont {Xi}~\bibnamefont {Dai}}, \ and\ \bibinfo {author}
  {\bibfnamefont {Zhong}\ \bibnamefont {Fang}},\ }\bibfield  {title} {\enquote
  {\bibinfo {title} {Chern semimetal and the quantized anomalous {H}all effect
  in ${\mathrm{hgcr}}_{2}{\mathrm{se}}_{4}$},}\ }\href {\doibase
  10.1103/PhysRevLett.107.186806} {\bibfield  {journal} {\bibinfo  {journal}
  {Phys. Rev. Lett.}\ }\textbf {\bibinfo {volume} {107}},\ \bibinfo {pages}
  {186806} (\bibinfo {year} {2011})}\BibitemShut {NoStop}%
\bibitem [{\citenamefont {Wan}\ \emph {et~al.}(2011)\citenamefont {Wan},
  \citenamefont {Turner}, \citenamefont {Vishwanath},\ and\ \citenamefont
  {Savrasov}}]{Wan2011}%
  \BibitemOpen
  \bibfield  {author} {\bibinfo {author} {\bibfnamefont {Xiangang}\
  \bibnamefont {Wan}}, \bibinfo {author} {\bibfnamefont {Ari~M.}\ \bibnamefont
  {Turner}}, \bibinfo {author} {\bibfnamefont {Ashvin}\ \bibnamefont
  {Vishwanath}}, \ and\ \bibinfo {author} {\bibfnamefont {Sergey~Y.}\
  \bibnamefont {Savrasov}},\ }\bibfield  {title} {\enquote {\bibinfo {title}
  {Topological semimetal and {F}ermi-arc surface states in the electronic
  structure of pyrochlore iridates},}\ }\href {\doibase
  10.1103/PhysRevB.83.205101} {\bibfield  {journal} {\bibinfo  {journal} {Phys.
  Rev. B}\ }\textbf {\bibinfo {volume} {83}},\ \bibinfo {pages} {205101}
  (\bibinfo {year} {2011})}\BibitemShut {NoStop}%
\bibitem [{\citenamefont {Liu}\ \emph {et~al.}(2014)\citenamefont {Liu},
  \citenamefont {Zhou}, \citenamefont {Zhang}, \citenamefont {Wang},
  \citenamefont {Weng}, \citenamefont {Prabhakaran}, \citenamefont {Mo},
  \citenamefont {Shen}, \citenamefont {Fang}, \citenamefont {Dai},
  \citenamefont {Hussain},\ and\ \citenamefont {Chen}}]{Liu2014}%
  \BibitemOpen
  \bibfield  {author} {\bibinfo {author} {\bibfnamefont {Z.~K.}\ \bibnamefont
  {Liu}}, \bibinfo {author} {\bibfnamefont {B.}~\bibnamefont {Zhou}}, \bibinfo
  {author} {\bibfnamefont {Y.}~\bibnamefont {Zhang}}, \bibinfo {author}
  {\bibfnamefont {Z.~J.}\ \bibnamefont {Wang}}, \bibinfo {author}
  {\bibfnamefont {H.~M.}\ \bibnamefont {Weng}}, \bibinfo {author}
  {\bibfnamefont {D.}~\bibnamefont {Prabhakaran}}, \bibinfo {author}
  {\bibfnamefont {S.-K.}\ \bibnamefont {Mo}}, \bibinfo {author} {\bibfnamefont
  {Z.~X.}\ \bibnamefont {Shen}}, \bibinfo {author} {\bibfnamefont
  {Z.}~\bibnamefont {Fang}}, \bibinfo {author} {\bibfnamefont {X.}~\bibnamefont
  {Dai}}, \bibinfo {author} {\bibfnamefont {Z.}~\bibnamefont {Hussain}}, \ and\
  \bibinfo {author} {\bibfnamefont {Y.~L.}\ \bibnamefont {Chen}},\ }\bibfield
  {title} {\enquote {\bibinfo {title} {Discovery of a three-dimensional
  topological {D}irac semimetal, {N}a$_3${B}i},}\ }\href {\doibase
  10.1126/science.1245085} {\bibfield  {journal} {\bibinfo  {journal}
  {Science}\ }\textbf {\bibinfo {volume} {343}},\ \bibinfo {pages} {864--867}
  (\bibinfo {year} {2014})}\BibitemShut {NoStop}%
\bibitem [{\citenamefont {Neupane}\ \emph {et~al.}(2014)\citenamefont
  {Neupane}, \citenamefont {Xu}, \citenamefont {Sankar}, \citenamefont
  {Alidoust}, \citenamefont {Bian}, \citenamefont {Liu}, \citenamefont
  {Belopolski}, \citenamefont {Chang}, \citenamefont {Jeng}, \citenamefont
  {Lin}, \citenamefont {Bansil}, \citenamefont {Chou},\ and\ \citenamefont
  {Hasan}}]{Neupane2014}%
  \BibitemOpen
  \bibfield  {author} {\bibinfo {author} {\bibfnamefont {Madhab}\ \bibnamefont
  {Neupane}}, \bibinfo {author} {\bibfnamefont {Su-Yang}\ \bibnamefont {Xu}},
  \bibinfo {author} {\bibfnamefont {Raman}\ \bibnamefont {Sankar}}, \bibinfo
  {author} {\bibfnamefont {Nasser}\ \bibnamefont {Alidoust}}, \bibinfo {author}
  {\bibfnamefont {Guang}\ \bibnamefont {Bian}}, \bibinfo {author}
  {\bibfnamefont {Chang}\ \bibnamefont {Liu}}, \bibinfo {author} {\bibfnamefont
  {Ilya}\ \bibnamefont {Belopolski}}, \bibinfo {author} {\bibfnamefont
  {Tay-Rong}\ \bibnamefont {Chang}}, \bibinfo {author} {\bibfnamefont
  {Horng-Tay}\ \bibnamefont {Jeng}}, \bibinfo {author} {\bibfnamefont {Hsin}\
  \bibnamefont {Lin}}, \bibinfo {author} {\bibfnamefont {Arun}\ \bibnamefont
  {Bansil}}, \bibinfo {author} {\bibfnamefont {Fangcheng}\ \bibnamefont
  {Chou}}, \ and\ \bibinfo {author} {\bibfnamefont {M.~Zahid}\ \bibnamefont
  {Hasan}},\ }\bibfield  {title} {\enquote {\bibinfo {title} {Observation of a
  three-dimensional topological {D}irac semimetal phase in high-mobility
  {C}d$_3${A}s$_2$},}\ }\href {http://dx.doi.org/10.1038/ncomms4786} {\bibfield
   {journal} {\bibinfo  {journal} {Nature Communications}\ }\textbf {\bibinfo
  {volume} {5}},\ \bibinfo {pages} {3786} (\bibinfo {year} {2014})}\BibitemShut
  {NoStop}%
\bibitem [{\citenamefont {Lv}\ \emph {et~al.}(2015)\citenamefont {Lv},
  \citenamefont {Weng}, \citenamefont {Fu}, \citenamefont {Wang}, \citenamefont
  {Miao}, \citenamefont {Ma}, \citenamefont {Richard}, \citenamefont {Huang},
  \citenamefont {Zhao}, \citenamefont {Chen}, \citenamefont {Fang},
  \citenamefont {Dai}, \citenamefont {Qian},\ and\ \citenamefont
  {Ding}}]{Lv2015}%
  \BibitemOpen
  \bibfield  {author} {\bibinfo {author} {\bibfnamefont {B.~Q.}\ \bibnamefont
  {Lv}}, \bibinfo {author} {\bibfnamefont {H.~M.}\ \bibnamefont {Weng}},
  \bibinfo {author} {\bibfnamefont {B.~B.}\ \bibnamefont {Fu}}, \bibinfo
  {author} {\bibfnamefont {X.~P.}\ \bibnamefont {Wang}}, \bibinfo {author}
  {\bibfnamefont {H.}~\bibnamefont {Miao}}, \bibinfo {author} {\bibfnamefont
  {J.}~\bibnamefont {Ma}}, \bibinfo {author} {\bibfnamefont {P.}~\bibnamefont
  {Richard}}, \bibinfo {author} {\bibfnamefont {X.~C.}\ \bibnamefont {Huang}},
  \bibinfo {author} {\bibfnamefont {L.~X.}\ \bibnamefont {Zhao}}, \bibinfo
  {author} {\bibfnamefont {G.~F.}\ \bibnamefont {Chen}}, \bibinfo {author}
  {\bibfnamefont {Z.}~\bibnamefont {Fang}}, \bibinfo {author} {\bibfnamefont
  {X.}~\bibnamefont {Dai}}, \bibinfo {author} {\bibfnamefont {T.}~\bibnamefont
  {Qian}}, \ and\ \bibinfo {author} {\bibfnamefont {H.}~\bibnamefont {Ding}},\
  }\bibfield  {title} {\enquote {\bibinfo {title} {Experimental discovery of
  {W}eyl semimetal {T}a{A}s},}\ }\href {\doibase 10.1103/PhysRevX.5.031013}
  {\bibfield  {journal} {\bibinfo  {journal} {Phys. Rev. X}\ }\textbf {\bibinfo
  {volume} {5}},\ \bibinfo {pages} {031013} (\bibinfo {year}
  {2015})}\BibitemShut {NoStop}%
\bibitem [{\citenamefont {Xu}\ \emph {et~al.}(2015)\citenamefont {Xu},
  \citenamefont {Belopolski}, \citenamefont {Alidoust}, \citenamefont
  {Neupane}, \citenamefont {Bian}, \citenamefont {Zhang}, \citenamefont
  {Sankar}, \citenamefont {Chang}, \citenamefont {Yuan}, \citenamefont {Lee},
  \citenamefont {Huang}, \citenamefont {Zheng}, \citenamefont {Ma},
  \citenamefont {Sanchez}, \citenamefont {Wang}, \citenamefont {Bansil},
  \citenamefont {Chou}, \citenamefont {Shibayev}, \citenamefont {Lin},
  \citenamefont {Jia},\ and\ \citenamefont {Hasan}}]{Xu2015}%
  \BibitemOpen
  \bibfield  {author} {\bibinfo {author} {\bibfnamefont {Su-Yang}\ \bibnamefont
  {Xu}}, \bibinfo {author} {\bibfnamefont {Ilya}\ \bibnamefont {Belopolski}},
  \bibinfo {author} {\bibfnamefont {Nasser}\ \bibnamefont {Alidoust}}, \bibinfo
  {author} {\bibfnamefont {Madhab}\ \bibnamefont {Neupane}}, \bibinfo {author}
  {\bibfnamefont {Guang}\ \bibnamefont {Bian}}, \bibinfo {author}
  {\bibfnamefont {Chenglong}\ \bibnamefont {Zhang}}, \bibinfo {author}
  {\bibfnamefont {Raman}\ \bibnamefont {Sankar}}, \bibinfo {author}
  {\bibfnamefont {Guoqing}\ \bibnamefont {Chang}}, \bibinfo {author}
  {\bibfnamefont {Zhujun}\ \bibnamefont {Yuan}}, \bibinfo {author}
  {\bibfnamefont {Chi-Cheng}\ \bibnamefont {Lee}}, \bibinfo {author}
  {\bibfnamefont {Shin-Ming}\ \bibnamefont {Huang}}, \bibinfo {author}
  {\bibfnamefont {Hao}\ \bibnamefont {Zheng}}, \bibinfo {author} {\bibfnamefont
  {Jie}\ \bibnamefont {Ma}}, \bibinfo {author} {\bibfnamefont {Daniel~S.}\
  \bibnamefont {Sanchez}}, \bibinfo {author} {\bibfnamefont {BaoKai}\
  \bibnamefont {Wang}}, \bibinfo {author} {\bibfnamefont {Arun}\ \bibnamefont
  {Bansil}}, \bibinfo {author} {\bibfnamefont {Fangcheng}\ \bibnamefont
  {Chou}}, \bibinfo {author} {\bibfnamefont {Pavel~P.}\ \bibnamefont
  {Shibayev}}, \bibinfo {author} {\bibfnamefont {Hsin}\ \bibnamefont {Lin}},
  \bibinfo {author} {\bibfnamefont {Shuang}\ \bibnamefont {Jia}}, \ and\
  \bibinfo {author} {\bibfnamefont {M.~Zahid}\ \bibnamefont {Hasan}},\
  }\bibfield  {title} {\enquote {\bibinfo {title} {Discovery of a {W}eyl
  fermion semimetal and topological {F}ermi arcs},}\ }\href {\doibase
  10.1126/science.aaa9297} {\bibfield  {journal} {\bibinfo  {journal}
  {Science}\ }\textbf {\bibinfo {volume} {349}},\ \bibinfo {pages} {613--617}
  (\bibinfo {year} {2015})}\BibitemShut {NoStop}%
\bibitem [{\citenamefont {Oka}\ and\ \citenamefont {Aoki}(2009)}]{Oka2009}%
  \BibitemOpen
  \bibfield  {author} {\bibinfo {author} {\bibfnamefont {Takashi}\ \bibnamefont
  {Oka}}\ and\ \bibinfo {author} {\bibfnamefont {Hideo}\ \bibnamefont {Aoki}},\
  }\bibfield  {title} {\enquote {\bibinfo {title} {Photovoltaic {H}all effect
  in graphene},}\ }\href {\doibase 10.1103/PhysRevB.79.081406} {\bibfield
  {journal} {\bibinfo  {journal} {Phys. Rev. B}\ }\textbf {\bibinfo {volume}
  {79}},\ \bibinfo {pages} {081406} (\bibinfo {year} {2009})}\BibitemShut
  {NoStop}%
\bibitem [{\citenamefont {Potter}\ \emph {et~al.}(2014)\citenamefont {Potter},
  \citenamefont {Kimchi},\ and\ \citenamefont {Vishwanath}}]{Potter2014}%
  \BibitemOpen
  \bibfield  {author} {\bibinfo {author} {\bibfnamefont {Andrew~C.}\
  \bibnamefont {Potter}}, \bibinfo {author} {\bibfnamefont {Itamar}\
  \bibnamefont {Kimchi}}, \ and\ \bibinfo {author} {\bibfnamefont {Ashvin}\
  \bibnamefont {Vishwanath}},\ }\bibfield  {title} {\enquote {\bibinfo {title}
  {Quantum oscillations from surface {F}ermi arcs in {W}eyl and {D}irac
  semimetals},}\ }\href {http://dx.doi.org/10.1038/ncomms6161} {\bibfield
  {journal} {\bibinfo  {journal} {Nature Communications}\ }\textbf {\bibinfo
  {volume} {5}},\ \bibinfo {pages} {5161} (\bibinfo {year} {2014})}\BibitemShut
  {NoStop}%
\bibitem [{\citenamefont {Moll}\ \emph {et~al.}(2016)\citenamefont {Moll},
  \citenamefont {Nair}, \citenamefont {Helm}, \citenamefont {Potter},
  \citenamefont {Kimchi}, \citenamefont {Vishwanath},\ and\ \citenamefont
  {Analytis}}]{Moll2016}%
  \BibitemOpen
  \bibfield  {author} {\bibinfo {author} {\bibfnamefont {Philip J.~W.}\
  \bibnamefont {Moll}}, \bibinfo {author} {\bibfnamefont {Nityan~L.}\
  \bibnamefont {Nair}}, \bibinfo {author} {\bibfnamefont {Toni}\ \bibnamefont
  {Helm}}, \bibinfo {author} {\bibfnamefont {Andrew~C.}\ \bibnamefont
  {Potter}}, \bibinfo {author} {\bibfnamefont {Itamar}\ \bibnamefont {Kimchi}},
  \bibinfo {author} {\bibfnamefont {Ashvin}\ \bibnamefont {Vishwanath}}, \ and\
  \bibinfo {author} {\bibfnamefont {James~G.}\ \bibnamefont {Analytis}},\
  }\bibfield  {title} {\enquote {\bibinfo {title} {Transport evidence for
  fermi-arc-mediated chirality transfer in the {D}irac semimetal
  {C}d$_3${A}s$_2$},}\ }\href {http://dx.doi.org/10.1038/nature18276}
  {\bibfield  {journal} {\bibinfo  {journal} {Nature}\ }\textbf {\bibinfo
  {volume} {535}},\ \bibinfo {pages} {266 -- 270} (\bibinfo {year}
  {2016})}\BibitemShut {NoStop}%
\bibitem [{\citenamefont {Taguchi}\ \emph {et~al.}(2016)\citenamefont
  {Taguchi}, \citenamefont {Imaeda}, \citenamefont {Sato},\ and\ \citenamefont
  {Tanaka}}]{Taguchi2016}%
  \BibitemOpen
  \bibfield  {author} {\bibinfo {author} {\bibfnamefont {Katsuhisa}\
  \bibnamefont {Taguchi}}, \bibinfo {author} {\bibfnamefont {Tatsushi}\
  \bibnamefont {Imaeda}}, \bibinfo {author} {\bibfnamefont {Masatoshi}\
  \bibnamefont {Sato}}, \ and\ \bibinfo {author} {\bibfnamefont {Yukio}\
  \bibnamefont {Tanaka}},\ }\bibfield  {title} {\enquote {\bibinfo {title}
  {Photovoltaic chiral magnetic effect in {W}eyl semimetals},}\ }\href
  {\doibase 10.1103/PhysRevB.93.201202} {\bibfield  {journal} {\bibinfo
  {journal} {Phys. Rev. B}\ }\textbf {\bibinfo {volume} {93}},\ \bibinfo
  {pages} {201202} (\bibinfo {year} {2016})}\BibitemShut {NoStop}%
\bibitem [{\citenamefont {Ebihara}\ \emph {et~al.}(2016)\citenamefont
  {Ebihara}, \citenamefont {Fukushima},\ and\ \citenamefont
  {Oka}}]{Ebihara2016}%
  \BibitemOpen
  \bibfield  {author} {\bibinfo {author} {\bibfnamefont {Shu}\ \bibnamefont
  {Ebihara}}, \bibinfo {author} {\bibfnamefont {Kenji}\ \bibnamefont
  {Fukushima}}, \ and\ \bibinfo {author} {\bibfnamefont {Takashi}\ \bibnamefont
  {Oka}},\ }\bibfield  {title} {\enquote {\bibinfo {title} {Chiral pumping
  effect induced by rotating electric fields},}\ }\href {\doibase
  10.1103/PhysRevB.93.155107} {\bibfield  {journal} {\bibinfo  {journal} {Phys.
  Rev. B}\ }\textbf {\bibinfo {volume} {93}},\ \bibinfo {pages} {155107}
  (\bibinfo {year} {2016})}\BibitemShut {NoStop}%
\bibitem [{\citenamefont {Chan}\ \emph {et~al.}(2016)\citenamefont {Chan},
  \citenamefont {Lee}, \citenamefont {Burch}, \citenamefont {Han},\ and\
  \citenamefont {Ran}}]{Chan2016}%
  \BibitemOpen
  \bibfield  {author} {\bibinfo {author} {\bibfnamefont {Ching-Kit}\
  \bibnamefont {Chan}}, \bibinfo {author} {\bibfnamefont {Patrick~A.}\
  \bibnamefont {Lee}}, \bibinfo {author} {\bibfnamefont {Kenneth~S.}\
  \bibnamefont {Burch}}, \bibinfo {author} {\bibfnamefont {Jung~Hoon}\
  \bibnamefont {Han}}, \ and\ \bibinfo {author} {\bibfnamefont {Ying}\
  \bibnamefont {Ran}},\ }\bibfield  {title} {\enquote {\bibinfo {title} {When
  chiral photons meet chiral fermions: Photoinduced anomalous hall effects in
  {W}eyl semimetals},}\ }\href {\doibase 10.1103/PhysRevLett.116.026805}
  {\bibfield  {journal} {\bibinfo  {journal} {Phys. Rev. Lett.}\ }\textbf
  {\bibinfo {volume} {116}},\ \bibinfo {pages} {026805} (\bibinfo {year}
  {2016})}\BibitemShut {NoStop}%
\bibitem [{\citenamefont {Ishizuka}\ \emph {et~al.}(2016)\citenamefont
  {Ishizuka}, \citenamefont {Hayata}, \citenamefont {Ueda},\ and\ \citenamefont
  {Nagaosa}}]{Ishizuka2016}%
  \BibitemOpen
  \bibfield  {author} {\bibinfo {author} {\bibfnamefont {Hiroaki}\ \bibnamefont
  {Ishizuka}}, \bibinfo {author} {\bibfnamefont {Tomoya}\ \bibnamefont
  {Hayata}}, \bibinfo {author} {\bibfnamefont {Masahito}\ \bibnamefont {Ueda}},
  \ and\ \bibinfo {author} {\bibfnamefont {Naoto}\ \bibnamefont {Nagaosa}},\
  }\bibfield  {title} {\enquote {\bibinfo {title} {Emergent electromagnetic
  induction and adiabatic charge pumping in noncentrosymmetric {W}eyl
  semimetals},}\ }\href {\doibase 10.1103/PhysRevLett.117.216601} {\bibfield
  {journal} {\bibinfo  {journal} {Phys. Rev. Lett.}\ }\textbf {\bibinfo
  {volume} {117}},\ \bibinfo {pages} {216601} (\bibinfo {year}
  {2016})}\BibitemShut {NoStop}%
\bibitem [{\citenamefont {Wu}\ \emph {et~al.}(2016)\citenamefont {Wu},
  \citenamefont {Patankar}, \citenamefont {Morimoto}, \citenamefont {Nair},
  \citenamefont {Thewalt}, \citenamefont {Little}, \citenamefont {Analytis},
  \citenamefont {Moore},\ and\ \citenamefont {Orenstein}}]{Wu2016}%
  \BibitemOpen
  \bibfield  {author} {\bibinfo {author} {\bibfnamefont {Liang}\ \bibnamefont
  {Wu}}, \bibinfo {author} {\bibfnamefont {S.}~\bibnamefont {Patankar}},
  \bibinfo {author} {\bibfnamefont {T.}~\bibnamefont {Morimoto}}, \bibinfo
  {author} {\bibfnamefont {N.~L.}\ \bibnamefont {Nair}}, \bibinfo {author}
  {\bibfnamefont {E.}~\bibnamefont {Thewalt}}, \bibinfo {author} {\bibfnamefont
  {A.}~\bibnamefont {Little}}, \bibinfo {author} {\bibfnamefont {J.~G.}\
  \bibnamefont {Analytis}}, \bibinfo {author} {\bibfnamefont {J.~E.}\
  \bibnamefont {Moore}}, \ and\ \bibinfo {author} {\bibfnamefont
  {J.}~\bibnamefont {Orenstein}},\ }\bibfield  {title} {\enquote {\bibinfo
  {title} {Giant anisotropic nonlinear optical response in transition metal
  monopnictide weyl semimetals},}\ }\href {http://dx.doi.org/10.1038/nphys3969}
  {\bibfield  {journal} {\bibinfo  {journal} {Nature Physics}\ }\textbf
  {\bibinfo {volume} {13}},\ \bibinfo {pages} {350} (\bibinfo {year}
  {2016})}\BibitemShut {NoStop}%
\bibitem [{\citenamefont {Ishizuka}\ \emph {et~al.}(2017)\citenamefont
  {Ishizuka}, \citenamefont {Hayata}, \citenamefont {Ueda},\ and\ \citenamefont
  {Nagaosa}}]{Ishizuka2017}%
  \BibitemOpen
  \bibfield  {author} {\bibinfo {author} {\bibfnamefont {Hiroaki}\ \bibnamefont
  {Ishizuka}}, \bibinfo {author} {\bibfnamefont {Tomoya}\ \bibnamefont
  {Hayata}}, \bibinfo {author} {\bibfnamefont {Masahito}\ \bibnamefont {Ueda}},
  \ and\ \bibinfo {author} {\bibfnamefont {Naoto}\ \bibnamefont {Nagaosa}},\
  }\bibfield  {title} {\enquote {\bibinfo {title} {Momentum-space
  electromagnetic induction in {W}eyl semimetals},}\ }\href {\doibase
  10.1103/PhysRevB.95.245211} {\bibfield  {journal} {\bibinfo  {journal} {Phys.
  Rev. B}\ }\textbf {\bibinfo {volume} {95}},\ \bibinfo {pages} {245211}
  (\bibinfo {year} {2017})}\BibitemShut {NoStop}%
\bibitem [{\citenamefont {Chan}\ \emph {et~al.}(2017)\citenamefont {Chan},
  \citenamefont {Lindner}, \citenamefont {Refael},\ and\ \citenamefont
  {Lee}}]{Chan2017}%
  \BibitemOpen
  \bibfield  {author} {\bibinfo {author} {\bibfnamefont {Ching-Kit}\
  \bibnamefont {Chan}}, \bibinfo {author} {\bibfnamefont {Netanel~H.}\
  \bibnamefont {Lindner}}, \bibinfo {author} {\bibfnamefont {Gil}\ \bibnamefont
  {Refael}}, \ and\ \bibinfo {author} {\bibfnamefont {Patrick~A.}\ \bibnamefont
  {Lee}},\ }\bibfield  {title} {\enquote {\bibinfo {title} {Photocurrents in
  {W}eyl semimetals},}\ }\href {\doibase 10.1103/PhysRevB.95.041104} {\bibfield
   {journal} {\bibinfo  {journal} {Phys. Rev. B}\ }\textbf {\bibinfo {volume}
  {95}},\ \bibinfo {pages} {041104} (\bibinfo {year} {2017})}\BibitemShut
  {NoStop}%
\bibitem [{\citenamefont {Ma}\ \emph {et~al.}(2017)\citenamefont {Ma},
  \citenamefont {Xu}, \citenamefont {Chan}, \citenamefont {Zhang},
  \citenamefont {Chang}, \citenamefont {Lin}, \citenamefont {Xie},
  \citenamefont {Palacios}, \citenamefont {Lin}, \citenamefont {Jia},
  \citenamefont {Lee}, \citenamefont {Jarillo-Herrero},\ and\ \citenamefont
  {Gedik}}]{Ma2017}%
  \BibitemOpen
  \bibfield  {author} {\bibinfo {author} {\bibfnamefont {Qiong}\ \bibnamefont
  {Ma}}, \bibinfo {author} {\bibfnamefont {Su-Yang}\ \bibnamefont {Xu}},
  \bibinfo {author} {\bibfnamefont {Ching-Kit}\ \bibnamefont {Chan}}, \bibinfo
  {author} {\bibfnamefont {Cheng-Long}\ \bibnamefont {Zhang}}, \bibinfo
  {author} {\bibfnamefont {Guoqing}\ \bibnamefont {Chang}}, \bibinfo {author}
  {\bibfnamefont {Yuxuan}\ \bibnamefont {Lin}}, \bibinfo {author}
  {\bibfnamefont {Weiwei}\ \bibnamefont {Xie}}, \bibinfo {author}
  {\bibfnamefont {Tomas}\ \bibnamefont {Palacios}}, \bibinfo {author}
  {\bibfnamefont {Hsin}\ \bibnamefont {Lin}}, \bibinfo {author} {\bibfnamefont
  {Shuang}\ \bibnamefont {Jia}}, \bibinfo {author} {\bibfnamefont {Patrick~A.}\
  \bibnamefont {Lee}}, \bibinfo {author} {\bibfnamefont {Pablo}\ \bibnamefont
  {Jarillo-Herrero}}, \ and\ \bibinfo {author} {\bibfnamefont {Nuh}\
  \bibnamefont {Gedik}},\ }\bibfield  {title} {\enquote {\bibinfo {title}
  {Direct optical detection of {W}eyl fermion chirality in a topological
  semimetal},}\ }\href {http://dx.doi.org/10.1038/nphys4146} {\bibfield
  {journal} {\bibinfo  {journal} {Nature Physics}\ }\textbf {\bibinfo {volume}
  {13}},\ \bibinfo {pages} {842 -- 847} (\bibinfo {year} {2017})}\BibitemShut
  {NoStop}%
\bibitem [{\citenamefont {de~Juan}\ \emph {et~al.}(2017)\citenamefont
  {de~Juan}, \citenamefont {Grushin}, \citenamefont {Morimoto},\ and\
  \citenamefont {Moore}}]{deJuan2017}%
  \BibitemOpen
  \bibfield  {author} {\bibinfo {author} {\bibfnamefont {Fernando}\
  \bibnamefont {de~Juan}}, \bibinfo {author} {\bibfnamefont {Adolfo~G.}\
  \bibnamefont {Grushin}}, \bibinfo {author} {\bibfnamefont {Takahiro}\
  \bibnamefont {Morimoto}}, \ and\ \bibinfo {author} {\bibfnamefont {Joel~E.}\
  \bibnamefont {Moore}},\ }\bibfield  {title} {\enquote {\bibinfo {title}
  {Quantized circular photogalvanic effect in {W}eyl semimetals},}\ }\href
  {http://dx.doi.org/10.1038/ncomms15995} {\bibfield  {journal} {\bibinfo
  {journal} {Nature Communications}\ }\textbf {\bibinfo {volume} {8}},\
  \bibinfo {pages} {15995} (\bibinfo {year} {2017})}\BibitemShut {NoStop}%
\bibitem [{\citenamefont {Osterhoudt}\ \emph {et~al.}()\citenamefont
  {Osterhoudt}, \citenamefont {Diebel}, \citenamefont {Yang}, \citenamefont
  {Stanco}, \citenamefont {Huang}, \citenamefont {Shen}, \citenamefont {Ni},
  \citenamefont {Moll}, \citenamefont {Ran},\ and\ \citenamefont
  {Burch}}]{Osterhoudt2017}%
  \BibitemOpen
  \bibfield  {author} {\bibinfo {author} {\bibfnamefont {Gavin~B.}\
  \bibnamefont {Osterhoudt}}, \bibinfo {author} {\bibfnamefont {Laura~K.}\
  \bibnamefont {Diebel}}, \bibinfo {author} {\bibfnamefont {Xu}~\bibnamefont
  {Yang}}, \bibinfo {author} {\bibfnamefont {John}\ \bibnamefont {Stanco}},
  \bibinfo {author} {\bibfnamefont {Xiangwei}\ \bibnamefont {Huang}}, \bibinfo
  {author} {\bibfnamefont {Bing}\ \bibnamefont {Shen}}, \bibinfo {author}
  {\bibfnamefont {Ni}~\bibnamefont {Ni}}, \bibinfo {author} {\bibfnamefont
  {Philip}\ \bibnamefont {Moll}}, \bibinfo {author} {\bibfnamefont {Ying}\
  \bibnamefont {Ran}}, \ and\ \bibinfo {author} {\bibfnamefont {Kenneth~S.}\
  \bibnamefont {Burch}},\ }\bibfield  {title} {\enquote {\bibinfo {title}
  {Colossal photovoltaic effect driven by the singular {B}erry curvature in a
  weyl semimetal},}\ }\href@noop {} {\bibinfo  {journal} {preprint}\ ,\
  \bibinfo {pages} {(arXiv:1712.04951)}}\BibitemShut {NoStop}%
\bibitem [{\citenamefont {Zhang}\ \emph
  {et~al.}(2018{\natexlab{a}})\citenamefont {Zhang}, \citenamefont {Sun},\ and\
  \citenamefont {Yan}}]{Zhang2018a}%
  \BibitemOpen
\bibfield  {journal} {  }\bibfield  {author} {\bibinfo {author} {\bibfnamefont
  {Yang}\ \bibnamefont {Zhang}}, \bibinfo {author} {\bibfnamefont {Yan}\
  \bibnamefont {Sun}}, \ and\ \bibinfo {author} {\bibfnamefont {Binghai}\
  \bibnamefont {Yan}},\ }\bibfield  {title} {\enquote {\bibinfo {title} {Berry
  curvature dipole in {W}eyl semimetal materials: An ab initio study},}\ }\href
  {\doibase 10.1103/PhysRevB.97.041101} {\bibfield  {journal} {\bibinfo
  {journal} {Phys. Rev. B}\ }\textbf {\bibinfo {volume} {97}},\ \bibinfo
  {pages} {041101} (\bibinfo {year} {2018}{\natexlab{a}})}\BibitemShut
  {NoStop}%
\bibitem [{\citenamefont {Zhang}\ \emph
  {et~al.}(2018{\natexlab{b}})\citenamefont {Zhang}, \citenamefont {Ishizuka},
  \citenamefont {van~den Brink}, \citenamefont {Felser}, \citenamefont {Yan},\
  and\ \citenamefont {Nagaosa}}]{Zhang2018b}%
  \BibitemOpen
  \bibfield  {author} {\bibinfo {author} {\bibfnamefont {Yang}\ \bibnamefont
  {Zhang}}, \bibinfo {author} {\bibfnamefont {Hiroaki}\ \bibnamefont
  {Ishizuka}}, \bibinfo {author} {\bibfnamefont {Jeroen}\ \bibnamefont {van~den
  Brink}}, \bibinfo {author} {\bibfnamefont {Claudia}\ \bibnamefont {Felser}},
  \bibinfo {author} {\bibfnamefont {Binghai}\ \bibnamefont {Yan}}, \ and\
  \bibinfo {author} {\bibfnamefont {Naoto}\ \bibnamefont {Nagaosa}},\
  }\bibfield  {title} {\enquote {\bibinfo {title} {Photogalvanic effect in
  {W}eyl semimetals from first principles},}\ }\href {\doibase
  10.1103/PhysRevB.97.241118} {\bibfield  {journal} {\bibinfo  {journal} {Phys.
  Rev. B}\ }\textbf {\bibinfo {volume} {97}},\ \bibinfo {pages} {241118}
  (\bibinfo {year} {2018}{\natexlab{b}})}\BibitemShut {NoStop}%
\bibitem [{\citenamefont {Liu}\ \emph {et~al.}(2013)\citenamefont {Liu},
  \citenamefont {Ye},\ and\ \citenamefont {Qi}}]{Liu2013}%
  \BibitemOpen
  \bibfield  {author} {\bibinfo {author} {\bibfnamefont {Chao-Xing}\
  \bibnamefont {Liu}}, \bibinfo {author} {\bibfnamefont {Peng}\ \bibnamefont
  {Ye}}, \ and\ \bibinfo {author} {\bibfnamefont {Xiao-Liang}\ \bibnamefont
  {Qi}},\ }\bibfield  {title} {\enquote {\bibinfo {title} {Chiral gauge field
  and axial anomaly in a {W}eyl semimetal},}\ }\href {\doibase
  10.1103/PhysRevB.87.235306} {\bibfield  {journal} {\bibinfo  {journal} {Phys.
  Rev. B}\ }\textbf {\bibinfo {volume} {87}},\ \bibinfo {pages} {235306}
  (\bibinfo {year} {2013})}\BibitemShut {NoStop}%
\bibitem [{\citenamefont {Guinea}\ \emph {et~al.}(2009)\citenamefont {Guinea},
  \citenamefont {Katsnelson},\ and\ \citenamefont {Geim}}]{Guinea2009}%
  \BibitemOpen
  \bibfield  {author} {\bibinfo {author} {\bibfnamefont {F.}~\bibnamefont
  {Guinea}}, \bibinfo {author} {\bibfnamefont {M.~I.}\ \bibnamefont
  {Katsnelson}}, \ and\ \bibinfo {author} {\bibfnamefont {A.~K.}\ \bibnamefont
  {Geim}},\ }\bibfield  {title} {\enquote {\bibinfo {title} {Energy gaps and a
  zero-field quantum hall effect in graphene by strain engineering},}\ }\href
  {http://dx.doi.org/10.1038/nphys1420} {\bibfield  {journal} {\bibinfo
  {journal} {Nature Physics}\ }\textbf {\bibinfo {volume} {6}},\ \bibinfo
  {pages} {30 -- 33} (\bibinfo {year} {2009})}\BibitemShut {NoStop}%
\bibitem [{\citenamefont {Levy}\ \emph {et~al.}(2010)\citenamefont {Levy},
  \citenamefont {Burke}, \citenamefont {Meaker}, \citenamefont {Panlasigui},
  \citenamefont {Zettl}, \citenamefont {Guinea}, \citenamefont {Neto},\ and\
  \citenamefont {Crommie}}]{Levy2010}%
  \BibitemOpen
  \bibfield  {author} {\bibinfo {author} {\bibfnamefont {N.}~\bibnamefont
  {Levy}}, \bibinfo {author} {\bibfnamefont {S.~A.}\ \bibnamefont {Burke}},
  \bibinfo {author} {\bibfnamefont {K.~L.}\ \bibnamefont {Meaker}}, \bibinfo
  {author} {\bibfnamefont {M.}~\bibnamefont {Panlasigui}}, \bibinfo {author}
  {\bibfnamefont {A.}~\bibnamefont {Zettl}}, \bibinfo {author} {\bibfnamefont
  {F.}~\bibnamefont {Guinea}}, \bibinfo {author} {\bibfnamefont {A.~H.~Castro}\
  \bibnamefont {Neto}}, \ and\ \bibinfo {author} {\bibfnamefont {M.~F.}\
  \bibnamefont {Crommie}},\ }\bibfield  {title} {\enquote {\bibinfo {title}
  {Strain-induced pseudo{\textendash}magnetic fields greater than 300 tesla in
  graphene nanobubbles},}\ }\href {\doibase 10.1126/science.1191700} {\bibfield
   {journal} {\bibinfo  {journal} {Science}\ }\textbf {\bibinfo {volume}
  {329}},\ \bibinfo {pages} {544--547} (\bibinfo {year} {2010})}\BibitemShut
  {NoStop}%
\bibitem [{\citenamefont {Chernodub}\ \emph {et~al.}(2014)\citenamefont
  {Chernodub}, \citenamefont {Cortijo}, \citenamefont {Grushin}, \citenamefont
  {Landsteiner},\ and\ \citenamefont {Vozmediano}}]{Chernodub2014}%
  \BibitemOpen
  \bibfield  {author} {\bibinfo {author} {\bibfnamefont {Maxim~N.}\
  \bibnamefont {Chernodub}}, \bibinfo {author} {\bibfnamefont {Alberto}\
  \bibnamefont {Cortijo}}, \bibinfo {author} {\bibfnamefont {Adolfo~G.}\
  \bibnamefont {Grushin}}, \bibinfo {author} {\bibfnamefont {Karl}\
  \bibnamefont {Landsteiner}}, \ and\ \bibinfo {author} {\bibfnamefont
  {Mar\'{\i}a A.~H.}\ \bibnamefont {Vozmediano}},\ }\bibfield  {title}
  {\enquote {\bibinfo {title} {Condensed matter realization of the axial
  magnetic effect},}\ }\href {\doibase 10.1103/PhysRevB.89.081407} {\bibfield
  {journal} {\bibinfo  {journal} {Phys. Rev. B}\ }\textbf {\bibinfo {volume}
  {89}},\ \bibinfo {pages} {081407} (\bibinfo {year} {2014})}\BibitemShut
  {NoStop}%
\bibitem [{\citenamefont {Cortijo}\ \emph {et~al.}(2015)\citenamefont
  {Cortijo}, \citenamefont {Ferreir\'os}, \citenamefont {Landsteiner},\ and\
  \citenamefont {Vozmediano}}]{Cortijo2015}%
  \BibitemOpen
  \bibfield  {author} {\bibinfo {author} {\bibfnamefont {Alberto}\ \bibnamefont
  {Cortijo}}, \bibinfo {author} {\bibfnamefont {Yago}\ \bibnamefont
  {Ferreir\'os}}, \bibinfo {author} {\bibfnamefont {Karl}\ \bibnamefont
  {Landsteiner}}, \ and\ \bibinfo {author} {\bibfnamefont {Mar\'{\i}a A.~H.}\
  \bibnamefont {Vozmediano}},\ }\bibfield  {title} {\enquote {\bibinfo {title}
  {Elastic gauge fields in {W}eyl semimetals},}\ }\href {\doibase
  10.1103/PhysRevLett.115.177202} {\bibfield  {journal} {\bibinfo  {journal}
  {Phys. Rev. Lett.}\ }\textbf {\bibinfo {volume} {115}},\ \bibinfo {pages}
  {177202} (\bibinfo {year} {2015})}\BibitemShut {NoStop}%
\bibitem [{\citenamefont {Pikulin}\ \emph {et~al.}(2016)\citenamefont
  {Pikulin}, \citenamefont {Chen},\ and\ \citenamefont {Franz}}]{Pikulin2016}%
  \BibitemOpen
  \bibfield  {author} {\bibinfo {author} {\bibfnamefont {D.~I.}\ \bibnamefont
  {Pikulin}}, \bibinfo {author} {\bibfnamefont {Anffany}\ \bibnamefont {Chen}},
  \ and\ \bibinfo {author} {\bibfnamefont {M.}~\bibnamefont {Franz}},\
  }\bibfield  {title} {\enquote {\bibinfo {title} {Chiral anomaly from
  strain-induced gauge fields in {D}irac and {W}eyl semimetals},}\ }\href
  {\doibase 10.1103/PhysRevX.6.041021} {\bibfield  {journal} {\bibinfo
  {journal} {Phys. Rev. X}\ }\textbf {\bibinfo {volume} {6}},\ \bibinfo {pages}
  {041021} (\bibinfo {year} {2016})}\BibitemShut {NoStop}%
\bibitem [{\citenamefont {Sumiyoshi}\ and\ \citenamefont
  {Fujimoto}(2016)}]{Sumiyoshi2016}%
  \BibitemOpen
  \bibfield  {author} {\bibinfo {author} {\bibfnamefont {Hiroaki}\ \bibnamefont
  {Sumiyoshi}}\ and\ \bibinfo {author} {\bibfnamefont {Satoshi}\ \bibnamefont
  {Fujimoto}},\ }\bibfield  {title} {\enquote {\bibinfo {title} {Torsional
  chiral magnetic effect in a {W}eyl semimetal with a topological defect},}\
  }\href {\doibase 10.1103/PhysRevLett.116.166601} {\bibfield  {journal}
  {\bibinfo  {journal} {Phys. Rev. Lett.}\ }\textbf {\bibinfo {volume} {116}},\
  \bibinfo {pages} {166601} (\bibinfo {year} {2016})}\BibitemShut {NoStop}%
\bibitem [{\citenamefont {Grushin}\ \emph {et~al.}(2016)\citenamefont
  {Grushin}, \citenamefont {Venderbos}, \citenamefont {Vishwanath},\ and\
  \citenamefont {Ilan}}]{Grushin2016}%
  \BibitemOpen
  \bibfield  {author} {\bibinfo {author} {\bibfnamefont {Adolfo~G.}\
  \bibnamefont {Grushin}}, \bibinfo {author} {\bibfnamefont {J\"orn W.~F.}\
  \bibnamefont {Venderbos}}, \bibinfo {author} {\bibfnamefont {Ashvin}\
  \bibnamefont {Vishwanath}}, \ and\ \bibinfo {author} {\bibfnamefont {Roni}\
  \bibnamefont {Ilan}},\ }\bibfield  {title} {\enquote {\bibinfo {title}
  {Inhomogeneous {W}eyl and {D}irac semimetals: Transport in axial magnetic
  fields and {F}ermi arc surface states from pseudo-{L}andau levels},}\ }\href
  {\doibase 10.1103/PhysRevX.6.041046} {\bibfield  {journal} {\bibinfo
  {journal} {Phys. Rev. X}\ }\textbf {\bibinfo {volume} {6}},\ \bibinfo {pages}
  {041046} (\bibinfo {year} {2016})}\BibitemShut {NoStop}%
\bibitem [{\citenamefont {Cortijo}\ \emph {et~al.}(2016)\citenamefont
  {Cortijo}, \citenamefont {Kharzeev}, \citenamefont {Landsteiner},\ and\
  \citenamefont {Vozmediano}}]{Cortijo2016}%
  \BibitemOpen
  \bibfield  {author} {\bibinfo {author} {\bibfnamefont {Alberto}\ \bibnamefont
  {Cortijo}}, \bibinfo {author} {\bibfnamefont {Dmitri}\ \bibnamefont
  {Kharzeev}}, \bibinfo {author} {\bibfnamefont {Karl}\ \bibnamefont
  {Landsteiner}}, \ and\ \bibinfo {author} {\bibfnamefont {Maria A.~H.}\
  \bibnamefont {Vozmediano}},\ }\bibfield  {title} {\enquote {\bibinfo {title}
  {Strain-induced chiral magnetic effect in {W}eyl semimetals},}\ }\href
  {\doibase 10.1103/PhysRevB.94.241405} {\bibfield  {journal} {\bibinfo
  {journal} {Phys. Rev. B}\ }\textbf {\bibinfo {volume} {94}},\ \bibinfo
  {pages} {241405} (\bibinfo {year} {2016})}\BibitemShut {NoStop}%
\bibitem [{\citenamefont {Gorbar}\ \emph {et~al.}(2017)\citenamefont {Gorbar},
  \citenamefont {Miransky}, \citenamefont {Shovkovy},\ and\ \citenamefont
  {Sukhachov}}]{Gorbar2017}%
  \BibitemOpen
  \bibfield  {author} {\bibinfo {author} {\bibfnamefont {E.~V.}\ \bibnamefont
  {Gorbar}}, \bibinfo {author} {\bibfnamefont {V.~A.}\ \bibnamefont
  {Miransky}}, \bibinfo {author} {\bibfnamefont {I.~A.}\ \bibnamefont
  {Shovkovy}}, \ and\ \bibinfo {author} {\bibfnamefont {P.~O.}\ \bibnamefont
  {Sukhachov}},\ }\bibfield  {title} {\enquote {\bibinfo {title}
  {Pseudomagnetic helicons},}\ }\href {\doibase 10.1103/PhysRevB.95.115422}
  {\bibfield  {journal} {\bibinfo  {journal} {Phys. Rev. B}\ }\textbf {\bibinfo
  {volume} {95}},\ \bibinfo {pages} {115422} (\bibinfo {year}
  {2017})}\BibitemShut {NoStop}%
\bibitem [{\citenamefont {Tchoumakov}\ \emph {et~al.}(2017)\citenamefont
  {Tchoumakov}, \citenamefont {Civelli},\ and\ \citenamefont
  {Goerbig}}]{Tchoumakov2017}%
  \BibitemOpen
  \bibfield  {author} {\bibinfo {author} {\bibfnamefont {Serguei}\ \bibnamefont
  {Tchoumakov}}, \bibinfo {author} {\bibfnamefont {Marcello}\ \bibnamefont
  {Civelli}}, \ and\ \bibinfo {author} {\bibfnamefont {Mark~O.}\ \bibnamefont
  {Goerbig}},\ }\bibfield  {title} {\enquote {\bibinfo {title} {Magnetic
  description of the {F}ermi arc in type-{I} and type-{II} {W}eyl
  semimetals},}\ }\href {\doibase 10.1103/PhysRevB.95.125306} {\bibfield
  {journal} {\bibinfo  {journal} {Phys. Rev. B}\ }\textbf {\bibinfo {volume}
  {95}},\ \bibinfo {pages} {125306} (\bibinfo {year} {2017})}\BibitemShut
  {NoStop}%
\bibitem [{\citenamefont {Kariyado}()}]{Kariyado2017}%
  \BibitemOpen
  \bibfield  {author} {\bibinfo {author} {\bibfnamefont {Toshikaze}\
  \bibnamefont {Kariyado}},\ }\bibfield  {title} {\enquote {\bibinfo {title}
  {Counting pseudo landau levels in spatially modulated {D}irac systems},}\
  }\href@noop {} {\bibinfo  {journal} {preprint (arXiv:1707.08601)}\
  }\BibitemShut {NoStop}%
\bibitem [{\citenamefont {Fukushima}\ \emph {et~al.}(2008)\citenamefont
  {Fukushima}, \citenamefont {Kharzeev},\ and\ \citenamefont
  {Warringa}}]{Fukushima2008}%
  \BibitemOpen
\bibfield  {journal} {  }\bibfield  {author} {\bibinfo {author} {\bibfnamefont
  {Kenji}\ \bibnamefont {Fukushima}}, \bibinfo {author} {\bibfnamefont
  {Dmitri~E.}\ \bibnamefont {Kharzeev}}, \ and\ \bibinfo {author}
  {\bibfnamefont {Harmen~J.}\ \bibnamefont {Warringa}},\ }\bibfield  {title}
  {\enquote {\bibinfo {title} {Chiral magnetic effect},}\ }\href {\doibase
  10.1103/PhysRevD.78.074033} {\bibfield  {journal} {\bibinfo  {journal} {Phys.
  Rev. D}\ }\textbf {\bibinfo {volume} {78}},\ \bibinfo {pages} {074033}
  (\bibinfo {year} {2008})}\BibitemShut {NoStop}%
\bibitem [{\citenamefont {Sekine}\ \emph {et~al.}(2017)\citenamefont {Sekine},
  \citenamefont {Culcer},\ and\ \citenamefont {MacDonald}}]{Sekine2017}%
  \BibitemOpen
  \bibfield  {author} {\bibinfo {author} {\bibfnamefont {Akihiko}\ \bibnamefont
  {Sekine}}, \bibinfo {author} {\bibfnamefont {Dimitrie}\ \bibnamefont
  {Culcer}}, \ and\ \bibinfo {author} {\bibfnamefont {Allan~H.}\ \bibnamefont
  {MacDonald}},\ }\bibfield  {title} {\enquote {\bibinfo {title} {Quantum
  kinetic theory of the chiral anomaly},}\ }\href {\doibase
  10.1103/PhysRevB.96.235134} {\bibfield  {journal} {\bibinfo  {journal} {Phys.
  Rev. B}\ }\textbf {\bibinfo {volume} {96}},\ \bibinfo {pages} {235134}
  (\bibinfo {year} {2017})}\BibitemShut {NoStop}%
\bibitem [{\citenamefont {Liang}\ \emph {et~al.}(2014)\citenamefont {Liang},
  \citenamefont {Gibson}, \citenamefont {Ali}, \citenamefont {Liu},
  \citenamefont {Cava},\ and\ \citenamefont {Ong}}]{Liang2014}%
  \BibitemOpen
  \bibfield  {author} {\bibinfo {author} {\bibfnamefont {Tian}\ \bibnamefont
  {Liang}}, \bibinfo {author} {\bibfnamefont {Quinn}\ \bibnamefont {Gibson}},
  \bibinfo {author} {\bibfnamefont {Mazhar~N.}\ \bibnamefont {Ali}}, \bibinfo
  {author} {\bibfnamefont {Minhao}\ \bibnamefont {Liu}}, \bibinfo {author}
  {\bibfnamefont {R.~J.}\ \bibnamefont {Cava}}, \ and\ \bibinfo {author}
  {\bibfnamefont {N.~P.}\ \bibnamefont {Ong}},\ }\bibfield  {title} {\enquote
  {\bibinfo {title} {Ultrahigh mobility and giant magnetoresistance in the
  {D}irac semimetal {C}d$_3${A}s$_2$},}\ }\href
  {http://dx.doi.org/10.1038/nmat4143} {\bibfield  {journal} {\bibinfo
  {journal} {Nature Materials}\ }\textbf {\bibinfo {volume} {14}},\ \bibinfo
  {pages} {280 -- 284} (\bibinfo {year} {2014})}\BibitemShut {NoStop}%
\bibitem [{\citenamefont {Huang}\ \emph {et~al.}(2015)\citenamefont {Huang},
  \citenamefont {Zhao}, \citenamefont {Long}, \citenamefont {Wang},
  \citenamefont {Chen}, \citenamefont {Yang}, \citenamefont {Liang},
  \citenamefont {Xue}, \citenamefont {Weng}, \citenamefont {Fang},
  \citenamefont {Dai},\ and\ \citenamefont {Chen}}]{Huang2015}%
  \BibitemOpen
  \bibfield  {author} {\bibinfo {author} {\bibfnamefont {Xiaochun}\
  \bibnamefont {Huang}}, \bibinfo {author} {\bibfnamefont {Lingxiao}\
  \bibnamefont {Zhao}}, \bibinfo {author} {\bibfnamefont {Yujia}\ \bibnamefont
  {Long}}, \bibinfo {author} {\bibfnamefont {Peipei}\ \bibnamefont {Wang}},
  \bibinfo {author} {\bibfnamefont {Dong}\ \bibnamefont {Chen}}, \bibinfo
  {author} {\bibfnamefont {Zhanhai}\ \bibnamefont {Yang}}, \bibinfo {author}
  {\bibfnamefont {Hui}\ \bibnamefont {Liang}}, \bibinfo {author} {\bibfnamefont
  {Mianqi}\ \bibnamefont {Xue}}, \bibinfo {author} {\bibfnamefont {Hongming}\
  \bibnamefont {Weng}}, \bibinfo {author} {\bibfnamefont {Zhong}\ \bibnamefont
  {Fang}}, \bibinfo {author} {\bibfnamefont {Xi}~\bibnamefont {Dai}}, \ and\
  \bibinfo {author} {\bibfnamefont {Genfu}\ \bibnamefont {Chen}},\ }\bibfield
  {title} {\enquote {\bibinfo {title} {Observation of the
  chiral-anomaly-induced negative magnetoresistance in 3d {W}eyl semimetal
  {T}a{A}s},}\ }\href {\doibase 10.1103/PhysRevX.5.031023} {\bibfield
  {journal} {\bibinfo  {journal} {Phys. Rev. X}\ }\textbf {\bibinfo {volume}
  {5}},\ \bibinfo {pages} {031023} (\bibinfo {year} {2015})}\BibitemShut
  {NoStop}%
\bibitem [{\citenamefont {Xiong}\ \emph {et~al.}(2015)\citenamefont {Xiong},
  \citenamefont {Kushwaha}, \citenamefont {Liang}, \citenamefont {Krizan},
  \citenamefont {Hirschberger}, \citenamefont {Wang}, \citenamefont {Cava},\
  and\ \citenamefont {Ong}}]{Xiong2015}%
  \BibitemOpen
  \bibfield  {author} {\bibinfo {author} {\bibfnamefont {Jun}\ \bibnamefont
  {Xiong}}, \bibinfo {author} {\bibfnamefont {Satya~K.}\ \bibnamefont
  {Kushwaha}}, \bibinfo {author} {\bibfnamefont {Tian}\ \bibnamefont {Liang}},
  \bibinfo {author} {\bibfnamefont {Jason~W.}\ \bibnamefont {Krizan}}, \bibinfo
  {author} {\bibfnamefont {Max}\ \bibnamefont {Hirschberger}}, \bibinfo
  {author} {\bibfnamefont {Wudi}\ \bibnamefont {Wang}}, \bibinfo {author}
  {\bibfnamefont {R.~J.}\ \bibnamefont {Cava}}, \ and\ \bibinfo {author}
  {\bibfnamefont {N.~P.}\ \bibnamefont {Ong}},\ }\bibfield  {title} {\enquote
  {\bibinfo {title} {Evidence for the chiral anomaly in the {D}irac semimetal
  {N}a$_3${B}i},}\ }\href {\doibase 10.1126/science.aac6089} {\bibfield
  {journal} {\bibinfo  {journal} {Science}\ }\textbf {\bibinfo {volume}
  {350}},\ \bibinfo {pages} {413--416} (\bibinfo {year} {2015})}\BibitemShut
  {NoStop}%
\bibitem [{\citenamefont {Li}\ \emph {et~al.}(2016{\natexlab{a}})\citenamefont
  {Li}, \citenamefont {Kharzeev}, \citenamefont {Zhang}, \citenamefont {Huang},
  \citenamefont {Pletikosito}, \citenamefont {Fedorov}, \citenamefont {Zhong},
  \citenamefont {Schneeloch}, \citenamefont {Gu},\ and\ \citenamefont
  {Valla}}]{Li2016b}%
  \BibitemOpen
  \bibfield  {author} {\bibinfo {author} {\bibfnamefont {Qiang}\ \bibnamefont
  {Li}}, \bibinfo {author} {\bibfnamefont {Dmitri~E.}\ \bibnamefont
  {Kharzeev}}, \bibinfo {author} {\bibfnamefont {Cheng}\ \bibnamefont {Zhang}},
  \bibinfo {author} {\bibfnamefont {Yuan}\ \bibnamefont {Huang}}, \bibinfo
  {author} {\bibfnamefont {I.}~\bibnamefont {Pletikosito}}, \bibinfo {author}
  {\bibfnamefont {A.~V.}\ \bibnamefont {Fedorov}}, \bibinfo {author}
  {\bibfnamefont {R.~D.}\ \bibnamefont {Zhong}}, \bibinfo {author}
  {\bibfnamefont {J.~A.}\ \bibnamefont {Schneeloch}}, \bibinfo {author}
  {\bibfnamefont {G.~D.}\ \bibnamefont {Gu}}, \ and\ \bibinfo {author}
  {\bibfnamefont {T.}~\bibnamefont {Valla}},\ }\bibfield  {title} {\enquote
  {\bibinfo {title} {Chiral magnetic effect in {Z}r{T}e$_5$},}\ }\href
  {http://dx.doi.org/10.1038/nphys3648} {\bibfield  {journal} {\bibinfo
  {journal} {Nature Physics}\ }\textbf {\bibinfo {volume} {12}},\ \bibinfo
  {pages} {550} (\bibinfo {year} {2016}{\natexlab{a}})}\BibitemShut {NoStop}%
\bibitem [{\citenamefont {Zhang}\ \emph {et~al.}(2016)\citenamefont {Zhang},
  \citenamefont {Xu}, \citenamefont {Belopolski}, \citenamefont {Yuan},
  \citenamefont {Lin}, \citenamefont {Tong}, \citenamefont {Bian},
  \citenamefont {Alidoust}, \citenamefont {Lee}, \citenamefont {Huang},
  \citenamefont {Chang}, \citenamefont {Chang}, \citenamefont {Hsu},
  \citenamefont {Jeng}, \citenamefont {Neupane}, \citenamefont {Sanchez},
  \citenamefont {Zheng}, \citenamefont {Wang}, \citenamefont {Lin},
  \citenamefont {Zhang}, \citenamefont {Lu}, \citenamefont {Shen},
  \citenamefont {Neupert}, \citenamefont {Zahid~Hasan},\ and\ \citenamefont
  {Jia}}]{Zhang2016}%
  \BibitemOpen
  \bibfield  {author} {\bibinfo {author} {\bibfnamefont {Cheng-Long}\
  \bibnamefont {Zhang}}, \bibinfo {author} {\bibfnamefont {Su-Yang}\
  \bibnamefont {Xu}}, \bibinfo {author} {\bibfnamefont {Ilya}\ \bibnamefont
  {Belopolski}}, \bibinfo {author} {\bibfnamefont {Zhujun}\ \bibnamefont
  {Yuan}}, \bibinfo {author} {\bibfnamefont {Ziquan}\ \bibnamefont {Lin}},
  \bibinfo {author} {\bibfnamefont {Bingbing}\ \bibnamefont {Tong}}, \bibinfo
  {author} {\bibfnamefont {Guang}\ \bibnamefont {Bian}}, \bibinfo {author}
  {\bibfnamefont {Nasser}\ \bibnamefont {Alidoust}}, \bibinfo {author}
  {\bibfnamefont {Chi-Cheng}\ \bibnamefont {Lee}}, \bibinfo {author}
  {\bibfnamefont {Shin-Ming}\ \bibnamefont {Huang}}, \bibinfo {author}
  {\bibfnamefont {Tay-Rong}\ \bibnamefont {Chang}}, \bibinfo {author}
  {\bibfnamefont {Guoqing}\ \bibnamefont {Chang}}, \bibinfo {author}
  {\bibfnamefont {Chuang-Han}\ \bibnamefont {Hsu}}, \bibinfo {author}
  {\bibfnamefont {Horng-Tay}\ \bibnamefont {Jeng}}, \bibinfo {author}
  {\bibfnamefont {Madhab}\ \bibnamefont {Neupane}}, \bibinfo {author}
  {\bibfnamefont {Daniel~S.}\ \bibnamefont {Sanchez}}, \bibinfo {author}
  {\bibfnamefont {Hao}\ \bibnamefont {Zheng}}, \bibinfo {author} {\bibfnamefont
  {Junfeng}\ \bibnamefont {Wang}}, \bibinfo {author} {\bibfnamefont {Hsin}\
  \bibnamefont {Lin}}, \bibinfo {author} {\bibfnamefont {Chi}\ \bibnamefont
  {Zhang}}, \bibinfo {author} {\bibfnamefont {Hai-Zhou}\ \bibnamefont {Lu}},
  \bibinfo {author} {\bibfnamefont {Shun-Qing}\ \bibnamefont {Shen}}, \bibinfo
  {author} {\bibfnamefont {Titus}\ \bibnamefont {Neupert}}, \bibinfo {author}
  {\bibfnamefont {M.}~\bibnamefont {Zahid~Hasan}}, \ and\ \bibinfo {author}
  {\bibfnamefont {Shuang}\ \bibnamefont {Jia}},\ }\bibfield  {title} {\enquote
  {\bibinfo {title} {Signatures of the {A}dler-{B}ell-{J}ackiw chiral anomaly
  in a {W}eyl fermion semimetal},}\ }\href
  {http://dx.doi.org/10.1038/ncomms10735} {\bibfield  {journal} {\bibinfo
  {journal} {Nature Communications}\ }\textbf {\bibinfo {volume} {7}},\
  \bibinfo {pages} {10735} (\bibinfo {year} {2016})}\BibitemShut {NoStop}%
\bibitem [{\citenamefont {Hirschberger}\ \emph {et~al.}(2016)\citenamefont
  {Hirschberger}, \citenamefont {Kushwaha}, \citenamefont {Wang}, \citenamefont
  {Gibson}, \citenamefont {Liang}, \citenamefont {Belvin}, \citenamefont
  {Bernevig}, \citenamefont {Cava},\ and\ \citenamefont
  {Ong}}]{Hirschberger2016}%
  \BibitemOpen
  \bibfield  {author} {\bibinfo {author} {\bibfnamefont {Max}\ \bibnamefont
  {Hirschberger}}, \bibinfo {author} {\bibfnamefont {Satya}\ \bibnamefont
  {Kushwaha}}, \bibinfo {author} {\bibfnamefont {Zhijun}\ \bibnamefont {Wang}},
  \bibinfo {author} {\bibfnamefont {Quinn}\ \bibnamefont {Gibson}}, \bibinfo
  {author} {\bibfnamefont {Sihang}\ \bibnamefont {Liang}}, \bibinfo {author}
  {\bibfnamefont {Carina}\ \bibnamefont {Belvin}}, \bibinfo {author}
  {\bibfnamefont {B.~A.}\ \bibnamefont {Bernevig}}, \bibinfo {author}
  {\bibfnamefont {R.~J.}\ \bibnamefont {Cava}}, \ and\ \bibinfo {author}
  {\bibfnamefont {N.~P.}\ \bibnamefont {Ong}},\ }\bibfield  {title} {\enquote
  {\bibinfo {title} {The chiral anomaly and thermopower of {W}eyl fermions in
  the half-{H}eusler {G}d{P}t{B}i},}\ }\href
  {http://dx.doi.org/10.1038/nmat4684} {\bibfield  {journal} {\bibinfo
  {journal} {Nature Materials}\ }\textbf {\bibinfo {volume} {15}},\ \bibinfo
  {pages} {1161 -- 1165} (\bibinfo {year} {2016})}\BibitemShut {NoStop}%
\bibitem [{\citenamefont {Kuroda}\ \emph {et~al.}(2017)\citenamefont {Kuroda},
  \citenamefont {Tomita}, \citenamefont {Suzuki}, \citenamefont {Bareille},
  \citenamefont {Nugroho}, \citenamefont {Goswami}, \citenamefont {Ochi},
  \citenamefont {Ikhlas}, \citenamefont {Nakayama}, \citenamefont {Akebi},
  \citenamefont {Noguchi}, \citenamefont {Ishii}, \citenamefont {Inami},
  \citenamefont {Ono}, \citenamefont {Kumigashira}, \citenamefont {Varykhalov},
  \citenamefont {Muro}, \citenamefont {Koretsune}, \citenamefont {Arita},
  \citenamefont {Shin}, \citenamefont {Kondo},\ and\ \citenamefont
  {Nakatsuji}}]{Kuroda2017}%
  \BibitemOpen
  \bibfield  {author} {\bibinfo {author} {\bibfnamefont {K.}~\bibnamefont
  {Kuroda}}, \bibinfo {author} {\bibfnamefont {T.}~\bibnamefont {Tomita}},
  \bibinfo {author} {\bibfnamefont {M.-T.}\ \bibnamefont {Suzuki}}, \bibinfo
  {author} {\bibfnamefont {C.}~\bibnamefont {Bareille}}, \bibinfo {author}
  {\bibfnamefont {A.~tuA}\ \bibnamefont {Nugroho}}, \bibinfo {author}
  {\bibfnamefont {P.}~\bibnamefont {Goswami}}, \bibinfo {author} {\bibfnamefont
  {M.}~\bibnamefont {Ochi}}, \bibinfo {author} {\bibfnamefont {M.}~\bibnamefont
  {Ikhlas}}, \bibinfo {author} {\bibfnamefont {M.}~\bibnamefont {Nakayama}},
  \bibinfo {author} {\bibfnamefont {S.}~\bibnamefont {Akebi}}, \bibinfo
  {author} {\bibfnamefont {R.}~\bibnamefont {Noguchi}}, \bibinfo {author}
  {\bibfnamefont {R.}~\bibnamefont {Ishii}}, \bibinfo {author} {\bibfnamefont
  {N.}~\bibnamefont {Inami}}, \bibinfo {author} {\bibfnamefont
  {K.}~\bibnamefont {Ono}}, \bibinfo {author} {\bibfnamefont {H.}~\bibnamefont
  {Kumigashira}}, \bibinfo {author} {\bibfnamefont {A.}~\bibnamefont
  {Varykhalov}}, \bibinfo {author} {\bibfnamefont {T.}~\bibnamefont {Muro}},
  \bibinfo {author} {\bibfnamefont {T.}~\bibnamefont {Koretsune}}, \bibinfo
  {author} {\bibfnamefont {R.}~\bibnamefont {Arita}}, \bibinfo {author}
  {\bibfnamefont {S.}~\bibnamefont {Shin}}, \bibinfo {author} {\bibfnamefont
  {Takeshi}\ \bibnamefont {Kondo}}, \ and\ \bibinfo {author} {\bibfnamefont
  {S.}~\bibnamefont {Nakatsuji}},\ }\bibfield  {title} {\enquote {\bibinfo
  {title} {Evidence for magnetic {W}eyl fermions in a correlated metal},}\
  }\href {http://dx.doi.org/10.1038/nmat4987} {\bibfield  {journal} {\bibinfo
  {journal} {Nature Materials}\ }\textbf {\bibinfo {volume} {16}},\ \bibinfo
  {pages} {1090} (\bibinfo {year} {2017})}\BibitemShut {NoStop}%
\bibitem [{\citenamefont {Niemann}\ \emph {et~al.}(2017)\citenamefont
  {Niemann}, \citenamefont {Gooth}, \citenamefont {Wu}, \citenamefont {Basler},
  \citenamefont {Sergelius}, \citenamefont {Huhne}, \citenamefont
  {Rellinghaus}, \citenamefont {Shekhar}, \citenamefont {Sus}, \citenamefont
  {Schmidt}, \citenamefont {Felser}, \citenamefont {Yan},\ and\ \citenamefont
  {Nielsch}}]{Niemann2017}%
  \BibitemOpen
  \bibfield  {author} {\bibinfo {author} {\bibfnamefont {Anna~Corinna}\
  \bibnamefont {Niemann}}, \bibinfo {author} {\bibfnamefont {Johannes}\
  \bibnamefont {Gooth}}, \bibinfo {author} {\bibfnamefont {Shu-Chun}\
  \bibnamefont {Wu}}, \bibinfo {author} {\bibfnamefont {Svenja}\ \bibnamefont
  {Basler}}, \bibinfo {author} {\bibfnamefont {Philip}\ \bibnamefont
  {Sergelius}}, \bibinfo {author} {\bibfnamefont {Ruben}\ \bibnamefont
  {Huhne}}, \bibinfo {author} {\bibfnamefont {Bernd}\ \bibnamefont
  {Rellinghaus}}, \bibinfo {author} {\bibfnamefont {Chandra}\ \bibnamefont
  {Shekhar}}, \bibinfo {author} {\bibfnamefont {Vicky}\ \bibnamefont {Sus}},
  \bibinfo {author} {\bibfnamefont {Marcus}\ \bibnamefont {Schmidt}}, \bibinfo
  {author} {\bibfnamefont {Claudia}\ \bibnamefont {Felser}}, \bibinfo {author}
  {\bibfnamefont {Binghai}\ \bibnamefont {Yan}}, \ and\ \bibinfo {author}
  {\bibfnamefont {Kornelius}\ \bibnamefont {Nielsch}},\ }\bibfield  {title}
  {\enquote {\bibinfo {title} {Chiral magnetoresistance in the {W}eyl semimetal
  {N}b{P}},}\ }\href {http://dx.doi.org/10.1038/srep43394} {\bibfield
  {journal} {\bibinfo  {journal} {Scientific Reports}\ }\textbf {\bibinfo
  {volume} {7}},\ \bibinfo {pages} {43394} (\bibinfo {year}
  {2017})}\BibitemShut {NoStop}%
\bibitem [{\citenamefont {Zhang}\ \emph {et~al.}(2017)\citenamefont {Zhang},
  \citenamefont {Zhang}, \citenamefont {Wang}, \citenamefont {Liu},
  \citenamefont {Chen}, \citenamefont {Lu}, \citenamefont {Liang},
  \citenamefont {Cao}, \citenamefont {Yuan}, \citenamefont {Tang},
  \citenamefont {Li}, \citenamefont {Zhou}, \citenamefont {Gu}, \citenamefont
  {Wu}, \citenamefont {Zou},\ and\ \citenamefont {Xiu}}]{Zhang2017}%
  \BibitemOpen
  \bibfield  {author} {\bibinfo {author} {\bibfnamefont {Cheng}\ \bibnamefont
  {Zhang}}, \bibinfo {author} {\bibfnamefont {Enze}\ \bibnamefont {Zhang}},
  \bibinfo {author} {\bibfnamefont {Weiyi}\ \bibnamefont {Wang}}, \bibinfo
  {author} {\bibfnamefont {Yanwen}\ \bibnamefont {Liu}}, \bibinfo {author}
  {\bibfnamefont {Zhi-Gang}\ \bibnamefont {Chen}}, \bibinfo {author}
  {\bibfnamefont {Shiheng}\ \bibnamefont {Lu}}, \bibinfo {author}
  {\bibfnamefont {Sihang}\ \bibnamefont {Liang}}, \bibinfo {author}
  {\bibfnamefont {Junzhi}\ \bibnamefont {Cao}}, \bibinfo {author}
  {\bibfnamefont {Xiang}\ \bibnamefont {Yuan}}, \bibinfo {author}
  {\bibfnamefont {Lei}\ \bibnamefont {Tang}}, \bibinfo {author} {\bibfnamefont
  {Qian}\ \bibnamefont {Li}}, \bibinfo {author} {\bibfnamefont {Chao}\
  \bibnamefont {Zhou}}, \bibinfo {author} {\bibfnamefont {Teng}\ \bibnamefont
  {Gu}}, \bibinfo {author} {\bibfnamefont {Yizheng}\ \bibnamefont {Wu}},
  \bibinfo {author} {\bibfnamefont {Jin}\ \bibnamefont {Zou}}, \ and\ \bibinfo
  {author} {\bibfnamefont {Faxian}\ \bibnamefont {Xiu}},\ }\bibfield  {title}
  {\enquote {\bibinfo {title} {Room-temperature chiral charge pumping in
  {D}irac semimetals},}\ }\href {http://dx.doi.org/10.1038/ncomms13741}
  {\bibfield  {journal} {\bibinfo  {journal} {Nature Communications}\ }\textbf
  {\bibinfo {volume} {8}},\ \bibinfo {pages} {13741} (\bibinfo {year}
  {2017})}\BibitemShut {NoStop}%
\bibitem [{\citenamefont {Liu}\ \emph {et~al.}(2018)\citenamefont {Liu},
  \citenamefont {Lin}, \citenamefont {Kushwaha}, \citenamefont {Cava},\ and\
  \citenamefont {Ong}}]{Liang2018}%
  \BibitemOpen
  \bibfield  {author} {\bibinfo {author} {\bibfnamefont {S.}~\bibnamefont
  {Liu}}, \bibinfo {author} {\bibfnamefont {J.}~\bibnamefont {Lin}}, \bibinfo
  {author} {\bibfnamefont {S.}~\bibnamefont {Kushwaha}}, \bibinfo {author}
  {\bibfnamefont {R.~J.}\ \bibnamefont {Cava}}, \ and\ \bibinfo {author}
  {\bibfnamefont {N.~P.}\ \bibnamefont {Ong}},\ }\bibfield  {title} {\enquote
  {\bibinfo {title} {Experimental tests of the chiral anomaly magnetoresistance
  in the {D}irac-{W}eyl semimetals {N}a$_3${B}i and {G}d{P}t{B}i},}\
  }\href@noop {} {\bibfield  {journal} {\bibinfo  {journal} {preprint
  (arXiv:1802.01544)}\ } (\bibinfo {year} {2018})}\BibitemShut {NoStop}%
\bibitem [{\citenamefont {Li}\ \emph {et~al.}(2016{\natexlab{b}})\citenamefont
  {Li}, \citenamefont {He}, \citenamefont {Lu}, \citenamefont {Zhang},
  \citenamefont {Liu}, \citenamefont {Ma}, \citenamefont {Fan}, \citenamefont
  {Shen},\ and\ \citenamefont {Wang}}]{Li2016}%
  \BibitemOpen
  \bibfield  {author} {\bibinfo {author} {\bibfnamefont {Hui}\ \bibnamefont
  {Li}}, \bibinfo {author} {\bibfnamefont {Hongtao}\ \bibnamefont {He}},
  \bibinfo {author} {\bibfnamefont {Hai-Zhou}\ \bibnamefont {Lu}}, \bibinfo
  {author} {\bibfnamefont {Huachen}\ \bibnamefont {Zhang}}, \bibinfo {author}
  {\bibfnamefont {Hongchao}\ \bibnamefont {Liu}}, \bibinfo {author}
  {\bibfnamefont {Rong}\ \bibnamefont {Ma}}, \bibinfo {author} {\bibfnamefont
  {Zhiyong}\ \bibnamefont {Fan}}, \bibinfo {author} {\bibfnamefont {Shun-Qing}\
  \bibnamefont {Shen}}, \ and\ \bibinfo {author} {\bibfnamefont {Jiannong}\
  \bibnamefont {Wang}},\ }\bibfield  {title} {\enquote {\bibinfo {title}
  {Negative magnetoresistance in {D}irac semimetal {C}d3{A}s2},}\ }\href
  {http://dx.doi.org/10.1038/ncomms10301} {\bibfield  {journal} {\bibinfo
  {journal} {Nature Communications}\ }\textbf {\bibinfo {volume} {7}},\
  \bibinfo {pages} {10301} (\bibinfo {year} {2016}{\natexlab{b}})}\BibitemShut
  {NoStop}%
\bibitem [{\citenamefont {Nishihaya}\ \emph {et~al.}(2018)\citenamefont
  {Nishihaya}, \citenamefont {Uchida}, \citenamefont {Nakazawa}, \citenamefont
  {Akiba}, \citenamefont {Kriener}, \citenamefont {Kozuka}, \citenamefont
  {Miyake}, \citenamefont {Taguchi}, \citenamefont {Tokunaga},\ and\
  \citenamefont {Kawasaki}}]{Nishihaya2018}%
  \BibitemOpen
  \bibfield  {author} {\bibinfo {author} {\bibfnamefont {S.}~\bibnamefont
  {Nishihaya}}, \bibinfo {author} {\bibfnamefont {M.}~\bibnamefont {Uchida}},
  \bibinfo {author} {\bibfnamefont {Y.}~\bibnamefont {Nakazawa}}, \bibinfo
  {author} {\bibfnamefont {K.}~\bibnamefont {Akiba}}, \bibinfo {author}
  {\bibfnamefont {M.}~\bibnamefont {Kriener}}, \bibinfo {author} {\bibfnamefont
  {Y.}~\bibnamefont {Kozuka}}, \bibinfo {author} {\bibfnamefont
  {A.}~\bibnamefont {Miyake}}, \bibinfo {author} {\bibfnamefont
  {Y.}~\bibnamefont {Taguchi}}, \bibinfo {author} {\bibfnamefont
  {M.}~\bibnamefont {Tokunaga}}, \ and\ \bibinfo {author} {\bibfnamefont
  {M.}~\bibnamefont {Kawasaki}},\ }\bibfield  {title} {\enquote {\bibinfo
  {title} {Negative magnetoresistance suppressed through a topological phase
  transition in
  $({\mathrm{cd}}_{1\ensuremath{-}x}{\mathrm{zn}}_{x}{)}_{3}{\mathrm{as}}_{2}$
  thin films},}\ }\href {\doibase 10.1103/PhysRevB.97.245103} {\bibfield
  {journal} {\bibinfo  {journal} {Phys. Rev. B}\ }\textbf {\bibinfo {volume}
  {97}},\ \bibinfo {pages} {245103} (\bibinfo {year} {2018})}\BibitemShut
  {NoStop}%
\bibitem [{\citenamefont {Sundaram}\ and\ \citenamefont
  {Niu}(1999)}]{Sundaram1999}%
  \BibitemOpen
  \bibfield  {author} {\bibinfo {author} {\bibfnamefont {Ganesh}\ \bibnamefont
  {Sundaram}}\ and\ \bibinfo {author} {\bibfnamefont {Qian}\ \bibnamefont
  {Niu}},\ }\bibfield  {title} {\enquote {\bibinfo {title} {Wave-packet
  dynamics in slowly perturbed crystals: Gradient corrections and {B}erry-phase
  effects},}\ }\href {\doibase 10.1103/PhysRevB.59.14915} {\bibfield  {journal}
  {\bibinfo  {journal} {Phys. Rev. B}\ }\textbf {\bibinfo {volume} {59}},\
  \bibinfo {pages} {14915--14925} (\bibinfo {year} {1999})}\BibitemShut
  {NoStop}%
\bibitem [{\citenamefont {Xiao}\ \emph {et~al.}(2010)\citenamefont {Xiao},
  \citenamefont {Chang},\ and\ \citenamefont {Niu}}]{Xiao2010}%
  \BibitemOpen
  \bibfield  {author} {\bibinfo {author} {\bibfnamefont {Di}~\bibnamefont
  {Xiao}}, \bibinfo {author} {\bibfnamefont {Ming-Che}\ \bibnamefont {Chang}},
  \ and\ \bibinfo {author} {\bibfnamefont {Qian}\ \bibnamefont {Niu}},\
  }\bibfield  {title} {\enquote {\bibinfo {title} {{B}erry phase effects on
  electronic properties},}\ }\href {\doibase 10.1103/RevModPhys.82.1959}
  {\bibfield  {journal} {\bibinfo  {journal} {Rev. Mod. Phys.}\ }\textbf
  {\bibinfo {volume} {82}},\ \bibinfo {pages} {1959--2007} (\bibinfo {year}
  {2010})}\BibitemShut {NoStop}%
\bibitem [{\citenamefont {Kim}\ \emph {et~al.}(2014)\citenamefont {Kim},
  \citenamefont {Kim},\ and\ \citenamefont {Sasaki}}]{Kim2014}%
  \BibitemOpen
  \bibfield  {author} {\bibinfo {author} {\bibfnamefont {Ki-Seok}\ \bibnamefont
  {Kim}}, \bibinfo {author} {\bibfnamefont {Heon-Jung}\ \bibnamefont {Kim}}, \
  and\ \bibinfo {author} {\bibfnamefont {M.}~\bibnamefont {Sasaki}},\
  }\bibfield  {title} {\enquote {\bibinfo {title} {Boltzmann equation approach
  to anomalous transport in a {W}eyl metal},}\ }\href {\doibase
  10.1103/PhysRevB.89.195137} {\bibfield  {journal} {\bibinfo  {journal} {Phys.
  Rev. B}\ }\textbf {\bibinfo {volume} {89}},\ \bibinfo {pages} {195137}
  (\bibinfo {year} {2014})}\BibitemShut {NoStop}%
\bibitem [{\citenamefont {Lundgren}\ \emph {et~al.}(2014)\citenamefont
  {Lundgren}, \citenamefont {Laurell},\ and\ \citenamefont
  {Fiete}}]{Lundgren2014}%
  \BibitemOpen
  \bibfield  {author} {\bibinfo {author} {\bibfnamefont {Rex}\ \bibnamefont
  {Lundgren}}, \bibinfo {author} {\bibfnamefont {Pontus}\ \bibnamefont
  {Laurell}}, \ and\ \bibinfo {author} {\bibfnamefont {Gregory~A.}\
  \bibnamefont {Fiete}},\ }\bibfield  {title} {\enquote {\bibinfo {title}
  {Thermoelectric properties of {W}eyl and {D}irac semimetals},}\ }\href
  {\doibase 10.1103/PhysRevB.90.165115} {\bibfield  {journal} {\bibinfo
  {journal} {Phys. Rev. B}\ }\textbf {\bibinfo {volume} {90}},\ \bibinfo
  {pages} {165115} (\bibinfo {year} {2014})}\BibitemShut {NoStop}%
\bibitem [{\citenamefont {Sharma}\ \emph {et~al.}(2016)\citenamefont {Sharma},
  \citenamefont {Goswami},\ and\ \citenamefont {Tewari}}]{Sharma2016}%
  \BibitemOpen
  \bibfield  {author} {\bibinfo {author} {\bibfnamefont {Girish}\ \bibnamefont
  {Sharma}}, \bibinfo {author} {\bibfnamefont {Pallab}\ \bibnamefont
  {Goswami}}, \ and\ \bibinfo {author} {\bibfnamefont {Sumanta}\ \bibnamefont
  {Tewari}},\ }\bibfield  {title} {\enquote {\bibinfo {title} {Nernst and
  magnetothermal conductivity in a lattice model of {W}eyl fermions},}\ }\href
  {\doibase 10.1103/PhysRevB.93.035116} {\bibfield  {journal} {\bibinfo
  {journal} {Phys. Rev. B}\ }\textbf {\bibinfo {volume} {93}},\ \bibinfo
  {pages} {035116} (\bibinfo {year} {2016})}\BibitemShut {NoStop}%
\bibitem [{\citenamefont {McCormick}\ \emph {et~al.}(2017)\citenamefont
  {McCormick}, \citenamefont {McKay},\ and\ \citenamefont
  {Trivedi}}]{McCormick2016}%
  \BibitemOpen
  \bibfield  {author} {\bibinfo {author} {\bibfnamefont {Timothy~M.}\
  \bibnamefont {McCormick}}, \bibinfo {author} {\bibfnamefont {Robert~C.}\
  \bibnamefont {McKay}}, \ and\ \bibinfo {author} {\bibfnamefont {Nandini}\
  \bibnamefont {Trivedi}},\ }\bibfield  {title} {\enquote {\bibinfo {title}
  {Semiclassical theory of anomalous transport in type-ii topological {W}eyl
  semimetals},}\ }\href {\doibase 10.1103/PhysRevB.96.235116} {\bibfield
  {journal} {\bibinfo  {journal} {Phys. Rev. B}\ }\textbf {\bibinfo {volume}
  {96}},\ \bibinfo {pages} {235116} (\bibinfo {year} {2017})}\BibitemShut
  {NoStop}%
\bibitem [{\citenamefont {Gao}\ \emph {et~al.}(2017)\citenamefont {Gao},
  \citenamefont {Yang},\ and\ \citenamefont {Niu}}]{Gao2017}%
  \BibitemOpen
  \bibfield  {author} {\bibinfo {author} {\bibfnamefont {Yang}\ \bibnamefont
  {Gao}}, \bibinfo {author} {\bibfnamefont {Shengyuan~A.}\ \bibnamefont
  {Yang}}, \ and\ \bibinfo {author} {\bibfnamefont {Qian}\ \bibnamefont
  {Niu}},\ }\bibfield  {title} {\enquote {\bibinfo {title} {Intrinsic relative
  magnetoconductivity of nonmagnetic metals},}\ }\href {\doibase
  10.1103/PhysRevB.95.165135} {\bibfield  {journal} {\bibinfo  {journal} {Phys.
  Rev. B}\ }\textbf {\bibinfo {volume} {95}},\ \bibinfo {pages} {165135}
  (\bibinfo {year} {2017})}\BibitemShut {NoStop}%
\bibitem [{\citenamefont {Sharma}\ \emph {et~al.}(2017)\citenamefont {Sharma},
  \citenamefont {Goswami},\ and\ \citenamefont {Tewari}}]{Sharma2017}%
  \BibitemOpen
  \bibfield  {author} {\bibinfo {author} {\bibfnamefont {Girish}\ \bibnamefont
  {Sharma}}, \bibinfo {author} {\bibfnamefont {Pallab}\ \bibnamefont
  {Goswami}}, \ and\ \bibinfo {author} {\bibfnamefont {Sumanta}\ \bibnamefont
  {Tewari}},\ }\bibfield  {title} {\enquote {\bibinfo {title} {Chiral anomaly
  and longitudinal magnetotransport in type-{II} {W}eyl semimetals},}\ }\href
  {\doibase 10.1103/PhysRevB.96.045112} {\bibfield  {journal} {\bibinfo
  {journal} {Phys. Rev. B}\ }\textbf {\bibinfo {volume} {96}},\ \bibinfo
  {pages} {045112} (\bibinfo {year} {2017})}\BibitemShut {NoStop}%
\bibitem [{\citenamefont {Wei}\ \emph {et~al.}(2018)\citenamefont {Wei},
  \citenamefont {Li}, \citenamefont {Qi},\ and\ \citenamefont
  {Feng}}]{Wei2018}%
  \BibitemOpen
  \bibfield  {author} {\bibinfo {author} {\bibfnamefont {Yi-Wen}\ \bibnamefont
  {Wei}}, \bibinfo {author} {\bibfnamefont {Chao-Kai}\ \bibnamefont {Li}},
  \bibinfo {author} {\bibfnamefont {Jingshan}\ \bibnamefont {Qi}}, \ and\
  \bibinfo {author} {\bibfnamefont {Ji}~\bibnamefont {Feng}},\ }\bibfield
  {title} {\enquote {\bibinfo {title} {Magnetoconductivity of type-{II} {W}eyl
  semimetals},}\ }\href {\doibase 10.1103/PhysRevB.97.205131} {\bibfield
  {journal} {\bibinfo  {journal} {Phys. Rev. B}\ }\textbf {\bibinfo {volume}
  {97}},\ \bibinfo {pages} {205131} (\bibinfo {year} {2018})}\BibitemShut
  {NoStop}%
\bibitem [{\citenamefont {Cortijo}(2016)}]{Cortijo2016b}%
  \BibitemOpen
  \bibfield  {author} {\bibinfo {author} {\bibfnamefont {Alberto}\ \bibnamefont
  {Cortijo}},\ }\bibfield  {title} {\enquote {\bibinfo {title} {Linear
  magnetochiral effect in {W}eyl semimetals},}\ }\href {\doibase
  10.1103/PhysRevB.94.241105} {\bibfield  {journal} {\bibinfo  {journal} {Phys.
  Rev. B}\ }\textbf {\bibinfo {volume} {94}},\ \bibinfo {pages} {241105}
  (\bibinfo {year} {2016})}\BibitemShut {NoStop}%
\bibitem [{\citenamefont {Yang}\ \emph {et~al.}(2014)\citenamefont {Yang},
  \citenamefont {Moon}, \citenamefont {Isobe},\ and\ \citenamefont
  {Nagaosa}}]{Yang2014}%
  \BibitemOpen
  \bibfield  {author} {\bibinfo {author} {\bibfnamefont {Bohm-Jung}\
  \bibnamefont {Yang}}, \bibinfo {author} {\bibfnamefont {Eun-Gook}\
  \bibnamefont {Moon}}, \bibinfo {author} {\bibfnamefont {Hiroki}\ \bibnamefont
  {Isobe}}, \ and\ \bibinfo {author} {\bibfnamefont {Naoto}\ \bibnamefont
  {Nagaosa}},\ }\bibfield  {title} {\enquote {\bibinfo {title} {Quantum
  criticality of topological phase transitions in three-dimensional interacting
  electronic systems},}\ }\href {http://dx.doi.org/10.1038/nphys3060}
  {\bibfield  {journal} {\bibinfo  {journal} {Nature Physics}\ }\textbf
  {\bibinfo {volume} {10}},\ \bibinfo {pages} {774 -- 778} (\bibinfo {year}
  {2014})}\BibitemShut {NoStop}%
\bibitem [{\citenamefont {Wang}\ \emph {et~al.}(2013)\citenamefont {Wang},
  \citenamefont {Weng}, \citenamefont {Wu}, \citenamefont {Dai},\ and\
  \citenamefont {Fang}}]{Wang2013}%
  \BibitemOpen
  \bibfield  {author} {\bibinfo {author} {\bibfnamefont {Zhijun}\ \bibnamefont
  {Wang}}, \bibinfo {author} {\bibfnamefont {Hongming}\ \bibnamefont {Weng}},
  \bibinfo {author} {\bibfnamefont {Quansheng}\ \bibnamefont {Wu}}, \bibinfo
  {author} {\bibfnamefont {Xi}~\bibnamefont {Dai}}, \ and\ \bibinfo {author}
  {\bibfnamefont {Zhong}\ \bibnamefont {Fang}},\ }\bibfield  {title} {\enquote
  {\bibinfo {title} {Three-dimensional {D}irac semimetal and quantum transport
  in {C}d${}_{3}${A}s${}_{2}$},}\ }\href {\doibase 10.1103/PhysRevB.88.125427}
  {\bibfield  {journal} {\bibinfo  {journal} {Phys. Rev. B}\ }\textbf {\bibinfo
  {volume} {88}},\ \bibinfo {pages} {125427} (\bibinfo {year}
  {2013})}\BibitemShut {NoStop}%
\bibitem [{\citenamefont {Witczak-Krempa}\ \emph {et~al.}(2013)\citenamefont
  {Witczak-Krempa}, \citenamefont {Go},\ and\ \citenamefont
  {Kim}}]{Witczak-Krempa2013}%
  \BibitemOpen
  \bibfield  {author} {\bibinfo {author} {\bibfnamefont {William}\ \bibnamefont
  {Witczak-Krempa}}, \bibinfo {author} {\bibfnamefont {Ara}\ \bibnamefont
  {Go}}, \ and\ \bibinfo {author} {\bibfnamefont {Yong~Baek}\ \bibnamefont
  {Kim}},\ }\bibfield  {title} {\enquote {\bibinfo {title} {Pyrochlore
  electrons under pressure, heat, and field: Shedding light on the iridates},}\
  }\href {\doibase 10.1103/PhysRevB.87.155101} {\bibfield  {journal} {\bibinfo
  {journal} {Phys. Rev. B}\ }\textbf {\bibinfo {volume} {87}},\ \bibinfo
  {pages} {155101} (\bibinfo {year} {2013})}\BibitemShut {NoStop}%
\bibitem [{\citenamefont {Matsuhira}\ \emph {et~al.}(2011)\citenamefont
  {Matsuhira}, \citenamefont {Wakeshima}, \citenamefont {Hinatsu},\ and\
  \citenamefont {Takagi}}]{Matsuhira2011}%
  \BibitemOpen
  \bibfield  {author} {\bibinfo {author} {\bibfnamefont {Kazuyuki}\
  \bibnamefont {Matsuhira}}, \bibinfo {author} {\bibfnamefont {Makoto}\
  \bibnamefont {Wakeshima}}, \bibinfo {author} {\bibfnamefont {Yukio}\
  \bibnamefont {Hinatsu}}, \ and\ \bibinfo {author} {\bibfnamefont {Seishi}\
  \bibnamefont {Takagi}},\ }\bibfield  {title} {\enquote {\bibinfo {title}
  {Metal-insulator transitions in pyrochlore oxides {L}n$_2${I}r$_2$o$_7$},}\
  }\href {\doibase 10.1143/JPSJ.80.094701} {\bibfield  {journal} {\bibinfo
  {journal} {Journal of the Physical Society of Japan}\ }\textbf {\bibinfo
  {volume} {80}},\ \bibinfo {pages} {094701} (\bibinfo {year}
  {2011})}\BibitemShut {NoStop}%
\bibitem [{\citenamefont {Ueda}\ \emph
  {et~al.}(2015{\natexlab{a}})\citenamefont {Ueda}, \citenamefont {Fujioka},
  \citenamefont {Yang}, \citenamefont {Shiogai}, \citenamefont {Tsukazaki},
  \citenamefont {Nakamura}, \citenamefont {Awaji}, \citenamefont {Nagaosa},\
  and\ \citenamefont {Tokura}}]{Ueda2015}%
  \BibitemOpen
  \bibfield  {author} {\bibinfo {author} {\bibfnamefont {K.}~\bibnamefont
  {Ueda}}, \bibinfo {author} {\bibfnamefont {J.}~\bibnamefont {Fujioka}},
  \bibinfo {author} {\bibfnamefont {B.-J.}\ \bibnamefont {Yang}}, \bibinfo
  {author} {\bibfnamefont {J.}~\bibnamefont {Shiogai}}, \bibinfo {author}
  {\bibfnamefont {A.}~\bibnamefont {Tsukazaki}}, \bibinfo {author}
  {\bibfnamefont {S.}~\bibnamefont {Nakamura}}, \bibinfo {author}
  {\bibfnamefont {S.}~\bibnamefont {Awaji}}, \bibinfo {author} {\bibfnamefont
  {N.}~\bibnamefont {Nagaosa}}, \ and\ \bibinfo {author} {\bibfnamefont
  {Y.}~\bibnamefont {Tokura}},\ }\bibfield  {title} {\enquote {\bibinfo {title}
  {Magnetic field-induced insulator-semimetal transition in a pyrochlore
  ${\mathrm{nd}}_{2}{\mathrm{ir}}_{2}{\mathrm{o}}_{7}$},}\ }\href {\doibase
  10.1103/PhysRevLett.115.056402} {\bibfield  {journal} {\bibinfo  {journal}
  {Phys. Rev. Lett.}\ }\textbf {\bibinfo {volume} {115}},\ \bibinfo {pages}
  {056402} (\bibinfo {year} {2015}{\natexlab{a}})}\BibitemShut {NoStop}%
\bibitem [{\citenamefont {Tian}\ \emph {et~al.}(2015)\citenamefont {Tian},
  \citenamefont {Kohama}, \citenamefont {Tomita}, \citenamefont {Ishizuka},
  \citenamefont {Hsieh}, \citenamefont {Ishikawa}, \citenamefont {Kindo},
  \citenamefont {Balents},\ and\ \citenamefont {Nakatsuji}}]{Tian2015}%
  \BibitemOpen
  \bibfield  {author} {\bibinfo {author} {\bibfnamefont {Zhaoming}\
  \bibnamefont {Tian}}, \bibinfo {author} {\bibfnamefont {Yoshimitsu}\
  \bibnamefont {Kohama}}, \bibinfo {author} {\bibfnamefont {Takahiro}\
  \bibnamefont {Tomita}}, \bibinfo {author} {\bibfnamefont {Hiroaki}\
  \bibnamefont {Ishizuka}}, \bibinfo {author} {\bibfnamefont {Timothy~H.}\
  \bibnamefont {Hsieh}}, \bibinfo {author} {\bibfnamefont {Jun~J.}\
  \bibnamefont {Ishikawa}}, \bibinfo {author} {\bibfnamefont {Koichi}\
  \bibnamefont {Kindo}}, \bibinfo {author} {\bibfnamefont {Leon}\ \bibnamefont
  {Balents}}, \ and\ \bibinfo {author} {\bibfnamefont {Satoru}\ \bibnamefont
  {Nakatsuji}},\ }\bibfield  {title} {\enquote {\bibinfo {title} {Field-induced
  quantum metal-insulator transition in the pyrochlore iridate
  {N}d$_2${I}r$_2${O}$_7$},}\ }\href {http://dx.doi.org/10.1038/nphys3567}
  {\bibfield  {journal} {\bibinfo  {journal} {Nature Physics}\ }\textbf
  {\bibinfo {volume} {12}},\ \bibinfo {pages} {134 -- 138} (\bibinfo {year}
  {2015})}\BibitemShut {NoStop}%
\bibitem [{\citenamefont {Ueda}\ \emph
  {et~al.}(2015{\natexlab{b}})\citenamefont {Ueda}, \citenamefont {Fujioka},
  \citenamefont {Terakura},\ and\ \citenamefont {Tokura}}]{Ueda2015b}%
  \BibitemOpen
  \bibfield  {author} {\bibinfo {author} {\bibfnamefont {K.}~\bibnamefont
  {Ueda}}, \bibinfo {author} {\bibfnamefont {J.}~\bibnamefont {Fujioka}},
  \bibinfo {author} {\bibfnamefont {C.}~\bibnamefont {Terakura}}, \ and\
  \bibinfo {author} {\bibfnamefont {Y.}~\bibnamefont {Tokura}},\ }\bibfield
  {title} {\enquote {\bibinfo {title} {Pressure and magnetic field effects on
  metal-insulator transitions of bulk and domain wall states in pyrochlore
  iridates},}\ }\href {\doibase 10.1103/PhysRevB.92.121110} {\bibfield
  {journal} {\bibinfo  {journal} {Phys. Rev. B}\ }\textbf {\bibinfo {volume}
  {92}},\ \bibinfo {pages} {121110} (\bibinfo {year}
  {2015}{\natexlab{b}})}\BibitemShut {NoStop}%
\end{thebibliography}
\end{document}